\documentclass[aps,prd,showpacs,superscriptaddress,preprintnumbers]{revtex4-2}

\usepackage{graphicx,float,wrapfig,subfigure}
\usepackage{amsfonts,amsmath,amssymb,amstext}
\usepackage{latexsym}
 \usepackage{bm}
\usepackage{color}
\usepackage[normalem]{ulem}
\usepackage{hyperref}

\newcommand{\tr}{\,\mbox{tr}}
\newcommand{\sign}{\,\mbox{sign}}
\definecolor{red}{rgb}{0.7,0,0}
\definecolor{blue}{rgb}{0,0,0.7}

\begin{document}

\title{Circularly polarized photon emission from magnetized chiral plasmas}
\date{November 27, 2024}

\author{Xinyang Wang}
\email{wangxy@aust.edu.cn}
\affiliation{Center for Fundamental Physics, School of Mechanics and Physics, Anhui University of Science and Technology, Huainan, Anhui 232001, People's Republic of China}

\author{Igor A. Shovkovy}
\email{igor.shovkovy@asu.edu}
\affiliation{College of Integrative Sciences and Arts, Arizona State University, Mesa, Arizona 85212, USA}
\affiliation{Department of Physics, Arizona State University, Tempe, Arizona 85287, USA}

\begin{abstract}
We investigate the emission of circularly polarized photons from a magnetized quark-gluon plasma with nonzero quark-number and chiral charge chemical potentials. These chemical potentials qualitatively influence the differential emission rates of circularly polarized photons. A nonzero net electric charge density, induced by quark-number chemical potentials, enhances the overall emission of one circular polarization over the other, while a nonzero chiral charge density introduces a spatial asymmetry in the emission with respect to reflection in the transverse plane. The signs of the electrical and chiral charge densities determine which circular polarization dominates overall and whether the emission preferentially aligns with or opposes the magnetic field. Based on these findings, we propose that polarized photon emission is a promising observable for characterizing the quark-gluon plasma produced in heavy-ion collisions.
\end{abstract}

\maketitle
\section{Introduction}

Over the past two decades, the study of exotic states of matter in extreme environments has become a central focus of modern nuclear physics. This interest is largely driven by successful experimental programs at the relativistic heavy-ion colliders in Brookhaven and CERN. The ultrarelativistic collisions produce quark-gluon plasma (QGP) at temperatures well above the quark deconfinement transition. By analyzing particle spectra and correlations, one uncovers valuable information about the fundamental properties of QGP.

High temperature and energy density are not the only extreme characteristics of the QGP produced by heavy-ion collisions. The system is, without exaggeration, rich in extremes. Since heavy ions carry electric charges, they induce extreme magnetic fields \cite{Skokov:2009qp,Voronyuk:2011jd,Deng:2012pc,Bloczynski:2012en,Guo:2019mgh}. Furthermore, large values of angular momentum in noncentral collisions can produce high vorticity in the resulting QGP fluid \cite{Jiang:2016woz,Deng:2016gyh}.  Many theoretical aspects of QGP associated with strong magnetic fields and high vorticity are routinely tested in experimental studies \cite{Abelev:2009ac,PhysRevC.81.054908,Selyuzhenkov:2011xq,Abelev:2012pa,Wang:2012qs,Ke:2012qb,Adamczyk:2013kcb,Adamczyk:2015eqo,Khachatryan:2016got,Acharya:2017fau,CMS:2017lrw,ALICE:2020siw,STAR:2021pwb,STAR:2022ahj}. (For reviews, see Refs.~\cite{Tuchin:2013ie,Kharzeev:2015znc,Huang:2015oca,Miransky:2015ava,Kharzeev:2024zzm}.)

One aspect we would like to discuss in more detail is the effect of chiral charge density on the electromagnetic probes of QGP. A nonzero chiral charge arises from an imbalance between left-handed and right-handed quarks in the plasma. At first glance, such a state may seem too exotic as it would break local parity symmetry. However, in a hot plasma, local bubbles of chiral plasma can be produced by topological configurations of gauge fields \cite{STAR:2023qyt}. This idea is not new, and it is taken seriously enough to motivate the challenging search for its observable signatures via the dipole and quadrupole chiral magnetic effects (CME) in multiple experiments \cite{Abelev:2009ac,PhysRevC.81.054908,Selyuzhenkov:2011xq,Abelev:2012pa,Wang:2012qs,Ke:2012qb,Adamczyk:2013kcb,Adamczyk:2015eqo,Khachatryan:2016got,Acharya:2017fau,CMS:2017lrw,ALICE:2020siw,STAR:2021pwb,STAR:2022ahj}. Unfortunately, the evidence supporting CME remains inconclusive.

This study was motivated by the proposition that there might be unexplored, better observables for detecting chiral plasma effects. Electromagnetic probes, such as photon and dilepton emissions, play a crucial role in revealing the properties of QGP in strong magnetic fields. These probes provide valuable information about temperature, magnetic fields, flow characteristics, and possibly much more. Naturally, one might ask how a nonzero chiral charge affects electromagnetic probes. Some attempts to address this question have been made before, see Ref.~\cite{Yee:2013qma,Mamo:2013jda,Mamo:2015xkw}. Here, we delve into more detail to illuminate the role of strong magnetic fields and propose new observables that may resolve the issue. While the task remains challenging, we believe it is not impossible. 

This paper is organized as follows. In Sec.~\ref{sec:Formalism}, we briefly present the formalism used for studying polarized photon emission from the plasma. The derivation of the polarization tensor for a plasma with nonzero quark number and chiral chemical potentials is outlined in Sec.~\ref{sec:Polarization}. Numerical results for the polarized photon emission are given in Sec.~\ref{sec:numerics}. In Sec.~\ref{Summary}, we discuss our main findings and summarize the results. Several technical derivations and auxiliary results are provided in the appendices at the end of the paper.

\section{Formalism}
\label{sec:Formalism}

\begin{figure}[t]
\centering
  \subfigure[]{\includegraphics[width=0.25\textwidth]{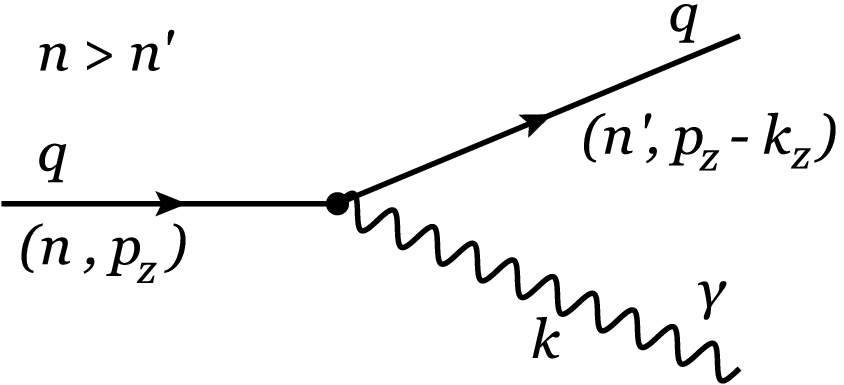}}
  \hspace{0.1\textwidth}
  \subfigure[]{\includegraphics[width=0.25\textwidth]{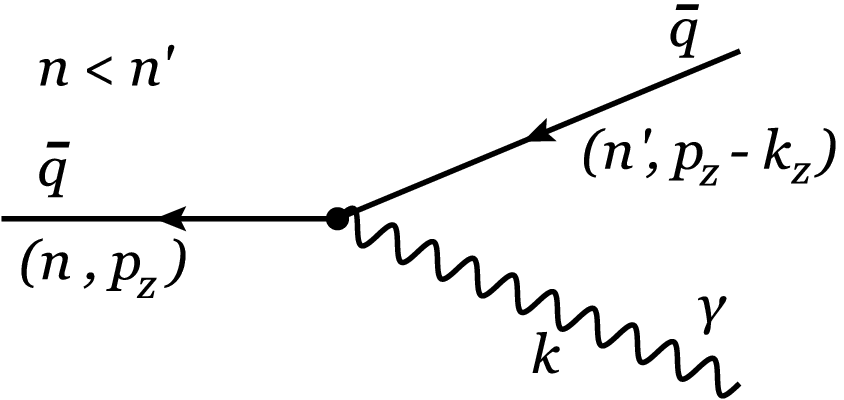}}
  \hspace{0.1\textwidth}
  \subfigure[]{\includegraphics[width=0.25\textwidth]{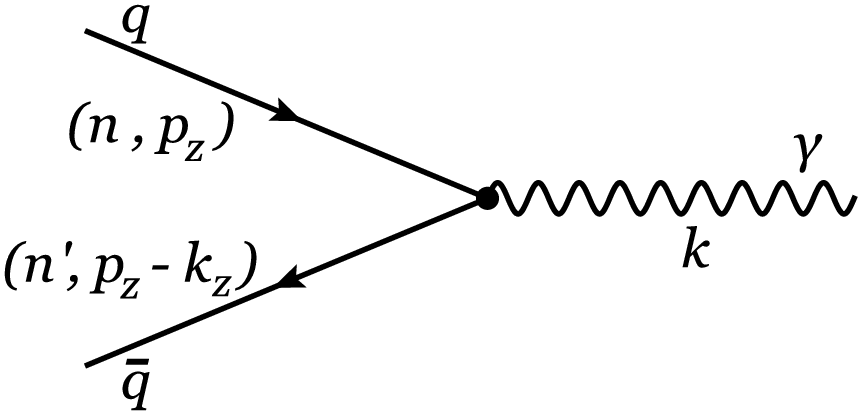}} 
  \caption{Three leading-order processes contributing to photon emission in a strongly magnetized QGP:
(a) $q\to q+\gamma$, (b) $\bar{q}\to \bar{q}+\gamma$, (c) $q+ \bar{q}\to \gamma$.}
\label{fig:processes}
\end{figure}

At leading order in the coupling, photon emission from a hot, strongly magnetized QGP is primarily driven by quark and antiquark splitting processes:
$q_n \to q_{n^\prime}+\gamma$ and $\bar{q}_n  \to \bar{q}_{n^\prime}+\gamma $, as well as by quark-antiquark annihilation, $q_n +\bar{q}_{n^\prime} \to \gamma$, shown in Fig.~\ref{fig:processes}. Here $q_n$ and $\bar{q}_{n^\prime}$ represent quark and antiquark states in the $n$th and $n^\prime$th Landau levels, respectively \cite{Wang:2020dsr}. These processes contribute at the linear order in the fine structure constant $\alpha=1/137$, but they are forbidden by energy-momentum conservation in the absence of a magnetic field. When $B=0$, the dominant contribution instead comes from gluon-mediated two-to-two processes: $q+g\rightarrow q+\gamma $, $\bar{q} +g \rightarrow \bar{q}+\gamma $, and $q + \bar{q}\rightarrow g+ \gamma$, where $g$ represents a gluon \cite{Kapusta:1991qp,Baier:1991em,Aurenche:1998nw,Steffen:2001pv,Arnold:2001ba,Arnold:2001ms,Ghiglieri:2013gia}.
The rates of these processes are of the order of $\alpha \alpha_s$, where $\alpha_s$ is the strong coupling constant. Thus, when the one-to-two and two-to-one processes become significant in the presence of a strong magnetic field, the two-to-two processes yield only subleading corrections. Here we assume that the QGP is at a sufficiently high temperature, ensuring that the running value of $\alpha_s$ remains small. However, even in this case, the interplay between the two types of processes becomes highly nontrivial as the magnetic field strength decreases.

To investigate the impact of a nonzero chiral charge density on photon emission from a strongly magnetized plasma, we will utilize the general framework developed in Refs.~\cite{Wang:2020dsr,Wang:2021ebh,Wang:2021eud}, where unpolarized photon emission was studied. Specifically, we will utilize a relationship between the emission rate and the imaginary part of the retarded photon polarization tensor \cite{Kapusta:2006pm}
\begin{equation}
{\cal R}_{\rm diff} \equiv k^0 \frac{d^3R}{dk_x dk_y dk_z}  
= - \frac{1}{(2\pi)^3} \frac{\mbox{Im}\left[g^{\mu \nu}\Pi_{\mu\nu}(k)\right]}{\exp\left(\frac{k_0}{T}\right)-1}.
\label{diff-rate}
\end{equation}
Assuming the background magnetic field is sufficiently strong, the imaginary part of the polarization tensor $\Pi_{\mu\nu}(k)$ is determined by the simplest one-loop diagram at leading order in the coupling \cite{Wang:2021ebh}. Since the branch cuts of the diagram are associated with intermediate two quark states going on the mass shell, $\mbox{Im}\left[\Pi_{\mu\nu}(k)\right]$ is proportional to the sum of squared amplitudes of the corresponding one-to-two and two-to-one processes, as depicted in Fig.~\ref{fig:processes}. For the QGP, this leading-order approximation holds when the temperature is sufficiently high and the magnetic field is strong enough, specifically when $T\gg \Lambda_{\rm QCD}$ and $|eB|\gtrsim \alpha_s T^2$. A similar approximation applies to a QED plasma when $|eB|\gtrsim \alpha T^2$.

To separate total photon emission into polarized components, we use the following completeness relation:
\begin{equation}
g^{\mu \nu} = \epsilon_{0}^\mu  \epsilon_{0}^{\nu*} - \epsilon_{+}^{\mu}  \epsilon_{+}^{\nu*} -  \epsilon_{-}^\mu  \epsilon_{-}^{\nu*} - \epsilon_{\parallel}^\mu  \epsilon_{\parallel}^{\nu*} ,
\label{decompose}
\end{equation}
where $\epsilon_{r}^{\mu}$ are photon polarization vectors. Here, we will use the same basis for $\epsilon_{r}^{\mu}$ as in Refs.~\cite{Mamo:2013jda,Mamo:2015xkw}, i.e., $\epsilon_{0}^\mu  = (1,0,0,0)$, $\epsilon_{\parallel}^\mu=(0,0,0,1)$,  $\epsilon_{\pm}^{\mu}=(0,1,\pm i, 0)/\sqrt{2}$. (A slightly different  decomposition was used in the study of dilepton production in Ref.~\cite{Hattori:2020htm}.)

This study focuses on the emission of circularly polarized photons, represented by the polarization vectors $\epsilon_{\pm}^{\mu}$. Before proceeding further, it is important to clarify the exact physical meaning of these photon states. For emission along (or opposite to) the direction of the magnetic field, these states correspond to the left-handed and right-handed circular polarizations, or equivalently, photons with positive and negative helicities, respectively. For photons emitted in other directions in the upper and lower hemispheres, they serve only as proxies for representing photons with given circular polarizations. For our qualitative discussion, we ignore the subtle difference between true circular polarization and its proxy defined by $\epsilon_{\pm}^{\mu}$. This is acceptable since, from symmetry considerations, both characterize closely related photon emission observables.

In this study, we focus on the emission of real photons into the vacuum outside of QGP. Since the mean free path of photons is much larger than the system size in heavy-ion collisions, intermediate modes in the plasma can be ignored. Once produced, these modes are projected onto vacuum states without any decoherence from scattering. As long as the corresponding set of states is complete, the emission rate of vacuum photons remains effectively unchanged. Similar arguments were used for dilepton production in Ref.~\cite{Wang:2022jxx}.

Note that the projection operators for circularly polarized photons take the following explicit form: 
\begin{equation}
\epsilon_{\pm}^{\mu}\epsilon_{\pm}^{\nu*} =  (-g_{\perp}^{\mu\nu} \pm i  \epsilon_{\perp }^{\mu\nu})/2 , 
\label{helicity-projectors}
\end{equation}
where $g_{\perp}^{\mu\nu} = g^{\mu\nu}-\delta^{\mu}_{0}\delta^{\nu}_{0}$ and $\epsilon_\perp^{\mu\nu}=\delta^{\mu}_{1}\delta^{\nu}_{2}-\delta^{\mu}_{2}\delta^{\nu }_{1}$. For completeness, let us note that the projectors for the other two states are $\epsilon_{0}^{\mu}\epsilon_{0}^{\nu*} =\mbox{diag}(1,0,0,0)$ and $\epsilon_{\parallel}^{\mu}\epsilon_{\parallel}^{\nu*} =\mbox{diag}(0,0,0,1)$. When these projectors are contracted with $\Pi_{\mu\nu}(k)$, as in Eq.~(\ref{diff-rate}), the photon polarization tensor naturally separates into the following four parts:
\begin{equation}
g^{\mu \nu}\Pi_{\mu\nu}(k)  \to  \Pi^{(0)}(k) + \Pi^{(+)}(k) + \Pi^{(-)}(k) +\Pi^{(\parallel)}(k) .\
\label{Pi-split}
\end{equation}
Then, the emission rate in Eq.~(\ref{diff-rate}) similarly splits into four parts. It is natural to interpret the contributions proportional to $\mbox{Im}\left[\Pi^{(+)}(k) \right]$ and $\mbox{Im}\left[\Pi^{(-)}(k) \right]$, 
\begin{eqnarray}
{\cal R}^{(+)}_{\rm diff} &\equiv& k^0 \frac{d^3R}{dk_x dk_y dk_z}  
= - \frac{1}{(2\pi)^3} \frac{\mbox{Im}\left[\Pi^{(+)}(k) \right]}{\exp\left(\frac{k_0}{T}\right)-1},
\label{diff-rate-plus} \\
{\cal R}^{(-)}_{\rm diff} &\equiv& k^0 \frac{d^3R}{dk_x dk_y dk_z}  
=  - \frac{1}{(2\pi)^3} \frac{\mbox{Im}\left[\Pi^{(-)}(k) \right]}{\exp\left(\frac{k_0}{T}\right)-1},
\label{diff-rate-minus}
\end{eqnarray}
as the emission rates for photons with the right-handed and the left-handed circular polarizations, respectively. It is important to emphasize, however, that such an interpretation is not entirely correct.
 
Recall that photons have only two physical degrees of freedom. Nevertheless, in the conventional quantum field theoretical description of a gauge theory like QED, it is convenient to introduce four states. Two of these states are unphysical. While they ensure the gauge invariance of the theory, they should not affect physical observables such as the photon emission rate. Thus, if all physical contributions were included in ${\cal R}^{(+)}_{\rm diff} $ and ${\cal R}^{(-)}_{\rm diff}$, the remaining parts would cancel out. However, this is not the case, partly due to the use of approximate rather than exact definitions of the polarization projectors in Eq.~(\ref{helicity-projectors}).

Interestingly, the extra contribution ${\cal R}^{\rm (extra)}_{\rm diff}$ to the rate, coming from $\Pi^{\rm (extra)}(k)\equiv \Pi^{(0)}(k)+\Pi^{(\parallel)}(k)$, is negative. At first glance, this might seem problematic since observable rates, which are proportional to the positive-definite probability of emitting physical photons, are expected to always be positive. In vacuum, there are only two physical photon states with transverse polarizations. Therefore, the additional contributions come from unphysical degrees of freedom that are necessary to preserve gauge invariance in the theoretical framework used. However, as one can verify, the total rate ${\cal R}_{\rm diff}$ remains positive overall. This suggests that the extra contribution ${\cal R}^{\rm (extra)}_{\rm diff}$ must be distributed between the two physical circular polarizations, thereby modifying the rate definitions in Eqs.~(\ref{diff-rate-plus}) and (\ref{diff-rate-minus}). For simplicity, we will continue using the unmodified definitions, which should be sufficient for approximately extracting parity-odd contributions to the rates for circularly polarized photons, albeit with a slight overestimation.

\section{Polarization tensor}
\label{sec:Polarization}

To calculate $\Pi_{\mu\nu}(k)$ in the QGP, we will assume that the plasma is made of the lightest up and down quarks. Their flavor-dependent quark charges are $e_f=eq_f$, where $f=u,d$, $q_{u} = 2/3$, $q_d =-1/3$,  and $e$ is the absolute value of the electron charge. When investigating the role of the chiral chemical potential $\mu_5$, it is convenient to use the chiral limit for quarks by setting their masses to zero. Noting that typical temperatures of a deconfined QGP are of the order of $200~\mbox{MeV}$ or higher and the current quark masses are only about $5~\mbox{MeV}$, the chiral limit is a very good approximation for the purposes of our study. 

Without loss of generality, we will take the background magnetic field $\mathbf{B}$ to be along the $+z$ direction. In the Dirac equation for the quarks, it will be described by a vector potential in the Landau gauge, i.e., $\mathbf{A}=(-B y,0,0)$. The corresponding quark Green's functions $G_f(u,u^\prime)$ take the form of a product of the so-called Schwinger phase $e^{i\Phi(u,u^\prime)}$ and a translation invariant function $\bar{G}_f(u-u^\prime)$, where $u=(t,x,y,z)$. Note, however, that  the Schwinger phase plays no role in the calculation of the polarization tensor $\Pi_{\mu\nu}(k)$ in the leading one-loop order approximation \cite{Wang:2021ebh}.

In a mixed coordinate-momentum space representation, the Fourier transform of the translation invariant part of Green's function can be conveniently given as a sum over the Landau levels~\cite{Miransky:2015ava}
\begin{equation}
\bar{G}_f(p_0, p_z ;\mathbf{r}_\perp) = i\frac{e^{-\mathbf{r}_\perp^2/(4\ell_f^2)}}{2\pi \ell_f^2}
\sum_{n=0}^{\infty}
\frac{\tilde{D}_{n}^{f}(p_0, p_z ;\mathbf{r}_\perp)}{\mathcal{M}-2n|e_fB|}, 
\label{GDn-alt}
\end{equation}
where the function in the numerator has the following explicit form:
\begin{equation}
\tilde{D}_{n}^{f}(p_0, p_z ;\mathbf{r}_\perp) = \left[(p_0+\mu)\gamma^0 -p_z\gamma^3 -\mu_5 \gamma^0 \gamma^5 \right]
\left[\mathcal{P}_{+}^fL_n\left(\frac{\mathbf{r}_{\perp}^2}{2\ell_f^{2}}\right)
+\mathcal{P}_{-}^fL_{n-1}\left(\frac{\mathbf{r}_{\perp}^2}{2\ell_f^{2}}\right)\right]
-\frac{i}{\ell_f^2}(\mathbf{r}_{\perp}\cdot\bm{\gamma}_{\perp}) 
 L_{n-1}^1\left(\frac{\mathbf{r}_{\perp}^2}{2\ell_f^{2}}\right) .
\end{equation}
Here, $L_{n}^\alpha(z)$ are the generalized Laguerre polynomials \cite{1980tisp.book.....G}, $\mathcal{P}_{\pm}^f\equiv \frac{1}{2} \left(1\pm i s_\perp \gamma^1\gamma^2\right)$ are the spin projectors, $\mathbf{r}_{\perp}=(x,y)$ is the position vector in the plane perpendicular to the field, and $\ell_f =1/\sqrt{|e_fB|}$ is the magnetic length. Here, we included nonzero quark number and chiral chemical potentials, denoted by $\mu$ and $\mu_5$, respectively. By definition, $s_\perp=\sign (e_f B)$ and $L_{-1}^\alpha(z) \equiv 0$. 

The function in the denominator of Eq.~(\ref{GDn-alt}) is a Dirac matrix. In the chiral limit, the corresponding fraction takes the following form:
\begin{equation}
\frac{1}{\mathcal{M}-2n|e_fB|} =   \frac{\mathcal{P}_5^{(-)}}{(p_{0}+\mu+\mu_5)^2-2n|e_fB|-p_z^2}+\frac{\mathcal{P}_5^{(+)}}{(p_{0}+\mu-\mu_5)^2-2n|e_fB|-p_z^2},
\end{equation}
where $\mathcal{P}_{5}^{(\pm)}=\frac{1}{2}(1\pm\gamma^5)$ are the chirality projectors. Note that we included both the quark-number and chiral-charge chemical potentials, $\mu$ and $\mu_5$, respectively. Both play an important role in polarized photon emission. For the sake of simplicity, we assume that $\mu$ and $\mu_5$ are same for both flavors. (The generalization to the case of flavor-dependent chemical potentials is straightforward and will be briefly discussed later.) 

By using the above Green's functions in the Landau-level representation, we derive the following formal expression for the polarization tensor \cite{Wang:2021ebh}:
\begin{equation}
\Pi^{\mu\nu}(i\Omega_m;\mathbf{k}) =  4\pi\sum_{f=u,d} N_c \alpha_f  T \sum_{k=-\infty}^{\infty} \int \frac{dp_z}{2\pi} 
\int d^2 \mathbf{r}_\perp e^{-i \mathbf{r}_\perp\cdot \mathbf{k}_\perp} 
\tr \left[ \gamma^\mu \bar{G}_f (i\omega_k, p_z ;\mathbf{r}_\perp)  
\gamma^\nu \bar{G}_f (i\omega_k-i\Omega_m, p_z-k_z; -\mathbf{r}_\perp)\right],
\label{Pi_Omega_k-alt}
\end{equation}
where we used the Matsubara finite-temperature formalism. By definition, the fermionic and bosonic Matsubara frequencies are $\omega_k= (2k+1)\pi T$ and $\Omega_m=2m \pi T$, respectively. Regarding other notations, $\alpha_f  =q_f^2 \alpha $ is the flavor-dependent coupling constant, and $N_c=3$ is the number of quark colors.

By substituting Green's function (\ref{GDn-alt}) into the definition for $\Pi^{\mu\nu}(i\Omega_m;\mathbf{k}) $, calculating the Dirac traces, and finally integrating over $r_{\perp}$, we derive the following expression for the individual components of the polarization tensor
\begin{equation}
\Pi^{(i)}(i\Omega_m;\mathbf{k})=  - \sum_{f=u,d}\frac{N_c \alpha_f T}{\pi \ell_f^4} \sum_{n,n^\prime = 0}^{\infty} \sum_{k=-\infty}^{\infty} \sum_{s=\pm 1}\int \frac{dp_z}{2\pi}\frac{\mathcal{J}_{(i)}^{f}(-s,-s)}{[(i\omega_k+\mu+s\mu_5)^2-E_{n,p_z,f}^2][(i\omega_k+\mu-i\Omega_m+s\mu_5)^2-E_{n^\prime,p_z-k_z,f}^2]  },
\end{equation}
where $i$ labels the photon polarizations (i.e., $i=0,+,-,\parallel$) and $E_{n,p_z,f}=\sqrt{p_z^2+2n|e_fB|}$ are Landau-level energies. The results for the Dirac traces are given in Appendix~\ref{app:Traces} and the explicit expressions for the functions $\mathcal{J}_{(i)}^{f}(-s,-s)$ are derived in Appendix~\ref{app:Integral-R-perp}.

After performing the Matsubara summation, we derive the following final expressions for the polarization components: 
\begin{equation}
\Pi^{(i)}(i\Omega_m;\mathbf{k})=-\sum_{f=u,d} \frac{N_c \alpha_f }{\pi \ell_f^4} \sum_{n,n^\prime = 0}^{\infty} \sum_{s,\eta,\lambda=\pm 1}\int \frac{dp_z}{2\pi}\frac{[n_F(E_{n,p_z,f}+\eta \mu+ \eta s \mu_5)-n_F(\lambda E_{n^\prime,p_z-k_z,f}+ \eta \mu+\eta s \mu_5)]\mathcal{F}^{(i)}_{s,f}(\xi, i\Omega_m)}{4\lambda E_{n,p_z,f}E_{n^\prime,p_z-k_z,f}\left[E_{n,p_z,f}-\lambda E_{n^\prime,p_z-k_z,f}+i \eta \Omega_m \right]},
\end{equation}
where $n_F(E) =1/\left[\exp(E/T)+1\right]$ is the Fermi-Dirac distribution function. The corresponding technical details as well as the explicit expressions for the functions $\mathcal{F}^{(i)}_{s,f}(\xi, i\Omega_m)$ are given in Appendix~\ref{app:ImaginaryPart}. 

After replacing $i\Omega_m \rightarrow k_0 + i\epsilon$, it is straightforward to extract the absorptive part of the individual polarization component functions,
\begin{eqnarray}
\label{Imp_f_s}
\mbox{Im}\Pi^{(i)} (k_{0};\mathbf{k}) &=&
 \sum_{f = u,d}\frac{N_c \alpha_f}{4\pi  \ell_f^4} \sum_{s,s^\prime=\pm1}\sum_{n>n^\prime}^{\infty} 
\frac{g(n,n^\prime) \, \theta\left[(k_{-}^f)^2+k_z^2-k_{0}^2\right] }{\sqrt{  \left[( k_{-}^f  )^2+k_z^2 - k_{0}^2 \right] \left[ (k_{+}^f )^2+k_z^2- k_{0}^2\right]  } } 
\left. \mathcal{F}^{(i)}_{s,f}(\xi,k_{0}) \right|_{p_z = p_{z,f}^{(s^\prime)} } 
\nonumber\\
&+& \sum_{f = u,d}\frac{N_c \alpha_f}{4\pi  \ell_f^4} \sum_{s,s^\prime=\pm 1} \sum_{n<n^\prime}^{\infty}  
\frac{g(n,n^\prime) \, \theta\left[(k_{-}^f)^2+k_z^2-k_{0}^2\right]  }{\sqrt{ \left[( k_{-}^f  )^2+k_z^2 - k_{0}^2 \right] \left[ (k_{+}^f )^2+k_z^2- k_{0}^2\right] } }
\left. \mathcal{F}^{(i)}_{s,f}(\xi,k_{0})\right|_{p_z = p_{z,f}^{(s^\prime)} } 
\nonumber\\
&-&\sum_{f = u,d}\frac{N_c \alpha_f}{4\pi \ell_f^4} \sum_{s,s^\prime=\pm 1} \sum_{n,n^\prime=0}^{\infty} 
\frac{g(n,n^\prime)\, \theta\left[k_{0}^2-k_z^2-(k_{+}^f)^2\right]}{\sqrt{ \left[k_{0}^2 - k_z^2-( k_{-}  ^f)^2 \right] \left[ k_{0}^2 - k_z^2-(k_{+}^f )^2\right]} }
\left. \mathcal{F}^{(i)}_{s,f}(\xi,k_{0}) \right|_{p_z = p_{z,f}^{(s^\prime)} } ,
\label{ImPi-final}
\end{eqnarray}
where  $\theta(x)$ is the Heaviside step function and $p_{z,f}^{(s^\prime)} $ are the solutions of the energy conservation equation, given in Eq.~(\ref{pz-solution}) in Appendix~\ref{app:ImaginaryPart}. In Eq.~(\ref{ImPi-final}), we also introduced two transverse momentum thresholds, 
\begin{equation}
\label{kpm}
k_{\pm}^{f} = \left|\sqrt{2n|e_fB|} \pm \sqrt{2n^{\prime}|e_fB|}\right| ,
\end{equation}
and the following function: 
\begin{equation}
g(n,n^\prime)=1-\sum_{s_1=\pm1}n_F \left(   \left[
\frac{k_{0}}{2}+s_1\frac{k_{0}(n-n^{\prime})|e_fB|}{k_{0}^2-k_z^2}+s_1 s^\prime \frac{|k_z|}{2}\sqrt{ \left(1-\frac{(k_{-}^{f})^2}{k_{0}^2-k_z^2} \right)\left( 1-\frac{(k_{+}^{f})^2}{k_{0}^2-k_z^2}\right)}\right]-s_1 \mu- s_1 s\mu_5 \right).\nonumber\\
\end{equation}
Note that the photon energy $k_{0}$ in the final expression satisfies the mass-shell condition, i.e., $k_{0} = \sqrt{k_\perp^2+k_z^2}$. 

As expected, the expression in Eq.~(\ref{ImPi-final}) closely resembles the result in Ref.~\cite{Wang:2021ebh}. It should be noted, however, that $g(n,n^\prime)$ depends on both fermion-number and chiral chemical potentials. The most important difference appears in the definition of functions $\mathcal{F}^{(i)}_{s,f}(\xi,\Omega)|_{p_z = p_{z,f}^{(s^\prime)} } $, which define individual polarization components labeled by $i=0,+,-,\parallel$. Their explicit expressions are given in Eqs.~(\ref{F-0-s-prime}) -- (\ref{F-par-s-prime}) in Appendix~\ref{app:ImaginaryPart}.

\section{Numerical results}
\label{sec:numerics}

\begin{figure}[t]
\centering
\includegraphics[height=0.3\textwidth]{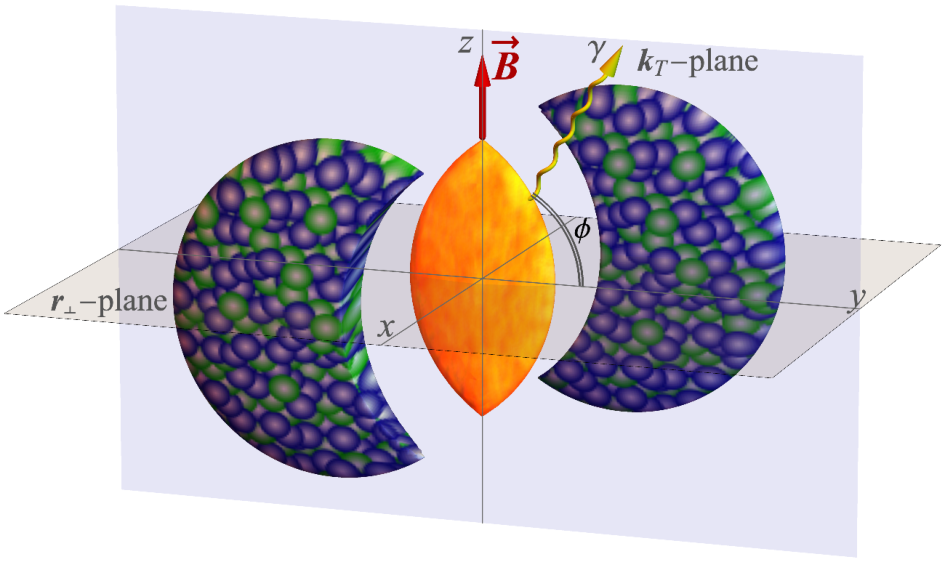}
\caption{The illustration of the coordinate system used, where the magnetic field $\mathbf{B}$ is perpendicular to the reaction plane. The reaction plane is parameterized by $\mathbf{r}_\perp=(x,y)$. The photon momentum $\mathbf{k}_T= k_T(0,\cos\phi,\sin\phi)$ lies in the transverse $yz$-plane, which is perpendicular to the beam.}
\label{fig:Illistration}
\end{figure}

Here, we present the main numerical results for the emission rates of circularly polarized photons. Considering potential applications to heavy-ion phenomenology, we will rewrite the differential rate using new variables: 
\begin{equation}
{\cal R}^{(\pm)}_{\rm diff} (k_T,\phi, y) =\frac{d^3 R^{(\pm)}}{k_T d k_T d\phi d y},
\end{equation}
where we introduced the transverse momentum $k_T$ (lying in the plane perpendicular to the ion beams), the azimuthal angle $\phi$, and the rapidity $y$, replacing the Cartesian components of the momentum, $k_x$, $k_y$, and $k_z$. By definition, $k_y = k_T \cos\phi$, $k_z = k_T \sin\phi$, and the rapidity $y = \frac{1}{2} \ln\frac{k_0+k_x}{k_0-k_x}$. For simplicity, below we will be presenting the results for zero rapidity (i.e., $k_x=0$). The illustration of the coordinate system used is shown in Fig.~\ref{fig:Illistration}.

In calculating the imaginary part of the polarization tensor, defined by Eq.~(\ref{ImPi-final}), we must sum over all Landau levels. For the numerical calculation, we will truncate the sum at a finite number of terms, specifically $n_{\rm max}=500$. This large number of Landau levels ensures high accuracy of the results across a wide range of model parameters.

For this study, we need to calculate the differential emission rate over the entire range of azimuthal angles, from $\phi_{\rm min}=-\pi/2$ to $\phi_{\rm max}= \pi/2$. To achieve high angular resolution, we use a relatively small step size, $\delta \phi  = \pi /600$. 

Our goal is to identify qualitative features of polarized photon emission in the presence of nonzero chemical potentials $\mu$ and $\mu_5$. Therefore, we limit our numerical analysis to a few representative values of the model parameters. Specifically, we consider two different values of the magnetic field strength, $|eB| = m_\pi^2$ and $|eB| = 5m_\pi^2$, and two different values of temperature, $T = 200~\mbox{MeV}$ and $T = 350~\mbox{MeV}$.

\subsection{Nonzero quark-number chemical potential $\mu$}
\label{subsec:mu}

Let us begin with the case of a nonzero quark-number chemical potential $\mu$. For our analysis, we will use a representative value of $\mu =200~\mbox{MeV}$. It is important to note that using the same chemical potential for both quark flavors implies that the plasma is positively charged overall, due to the up quark's charge being twice as large in magnitude compared to the down quark's charge. This analysis can be easily generalized to cases with different chemical potentials for the up and down quarks. As we argue below, it is the sign of the charge carriers in the plasma that determines the dominance of one circular polarization over the other in photon emission.

\begin{figure}[t]
\centering
{\includegraphics[width=0.44\textwidth]{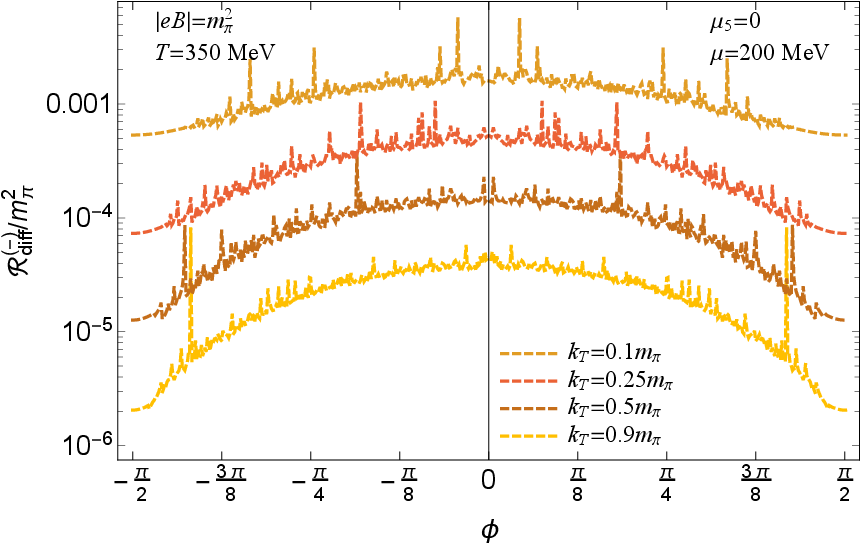}}
  \hspace{0.05\textwidth}
{\includegraphics[width=0.44\textwidth]{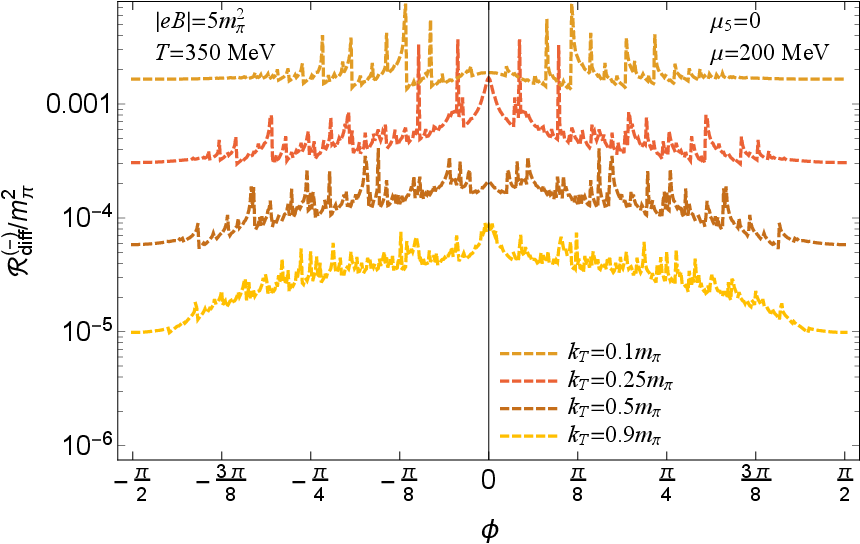}} \\[3mm]
{\includegraphics[width=0.44\textwidth]{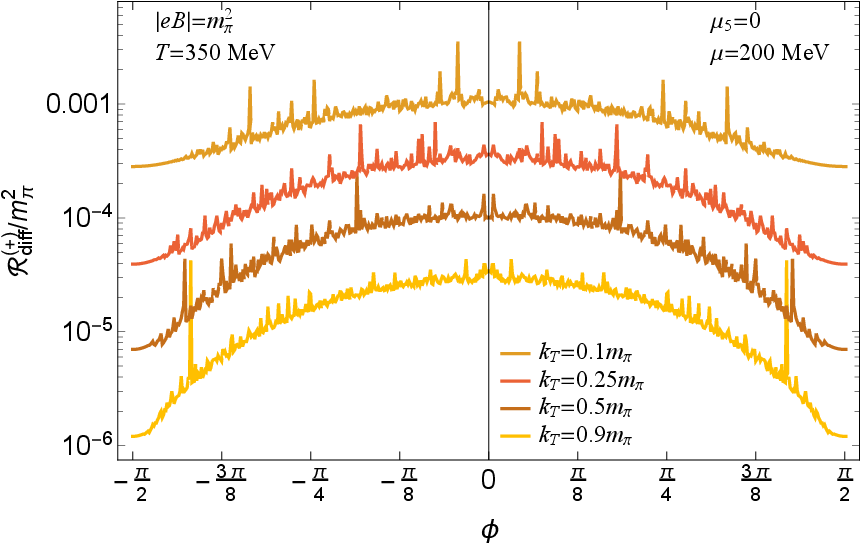}} 
  \hspace{0.05\textwidth}
{\includegraphics[width=0.44\textwidth]{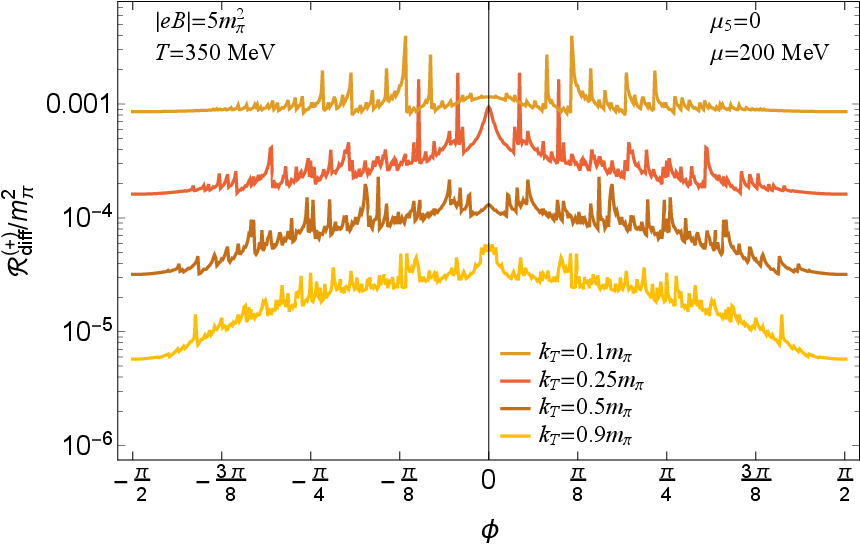}}
\caption{The angular dependence of left-handed (top) and right-handed (bottom) circularly polarized photon emission for two magnetic fields,  $|eB| = m_\pi^2$ (left) and $|eB| = 5m_\pi^2$ (right). The plasma temperature is $T = 350~\mbox{MeV}$.}
\label{fig:rates350-mu}
\end{figure}

Typical numerical results for the emission rates of left-handed and right-handed circularly polarized photons are presented in Fig.~\ref{fig:rates350-mu}. The data is shown for $T = 350~\mbox{MeV}$ and two different values of the magnetic field, $|eB| = m_\pi^2$ and $|eB| = 5m_\pi^2$, in the left and right panels, respectively. We verified that the qualitative features at  $T = 200~\mbox{MeV}$ are similar, though the rates are overall lower.

Each panel in Fig.~\ref{fig:rates350-mu} shows the differential rates for four different fixed values of the transverse momentum, $k_T=0.1 m_\pi, ~ 0.25 m_\pi, ~ 0.5 m_\pi , ~ 0.9 m_\pi$. Note that the transverse momentum also coincides with the photon energy since we set the rapidity to zero, $y=0$. The two top and two bottom panels display the results for left-handed and right-handed circular polarizations, respectively. As seen from the figure, the rates for both polarizations are symmetric under changing $\phi \to -\phi$, which is equivalent to reflection in the transverse plane. 

As we see, the differential rates exhibit nonsmooth behavior as a function of the angular coordinate $\phi$. They are characterized by numerous spikes resulting from the Landau-level quantization of quark states. This prominent feature was discussed in detail in Ref.~\cite{Wang:2021ebh}. Strictly speaking, the emergence of threshold spikes is an artifact of the approximation that neglects the self-energy of quarks. All threshold singularities would smooth out if the quark damping rates, $\Gamma_n(p_z)$, were considered. Notably, these damping rates were recently calculated in the Landau-level representation in Ref.~\cite{Ghosh:2024hbf}. Assuming small $\Gamma_n(p_z)$, the corresponding modifications occur only within narrow energy windows near the thresholds. Overall, the qualitative effect of nonzero $\Gamma_n(p_z)$ is expected to result in localized smoothing while preserving the global features of the rates \cite{Wang:2021ebh}. Thus, for the purposes of the current study, the inclusion of quark damping rates is not critical.

\begin{figure}[t]
\centering
{\includegraphics[width=0.44\textwidth]{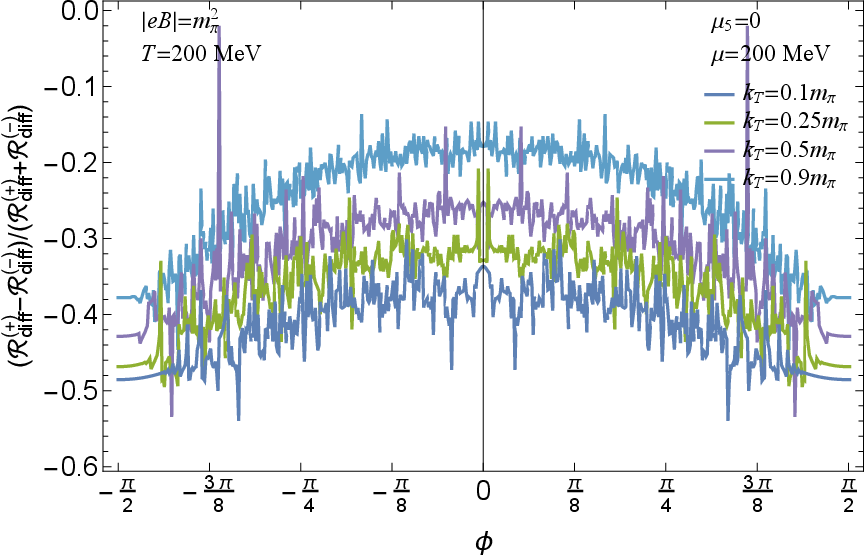}}
  \hspace{0.05\textwidth}
{\includegraphics[width=0.44\textwidth]{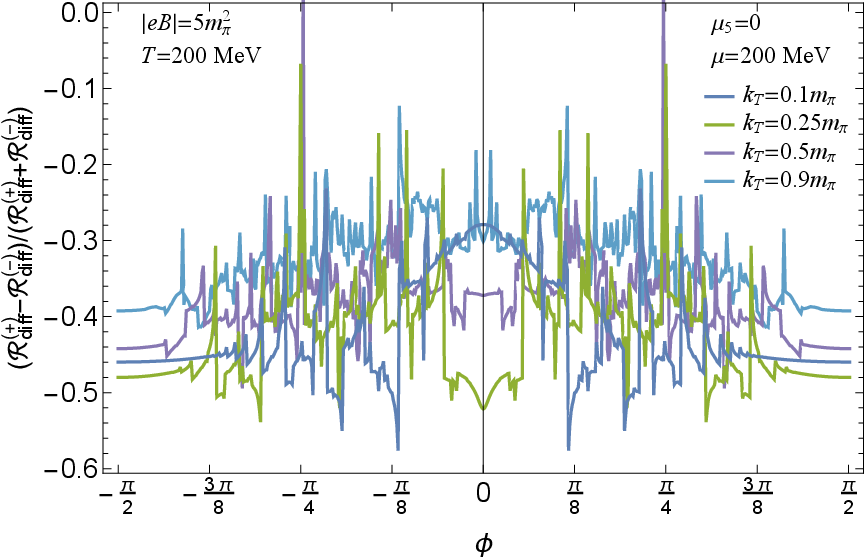}}\\
{\includegraphics[width=0.44\textwidth]{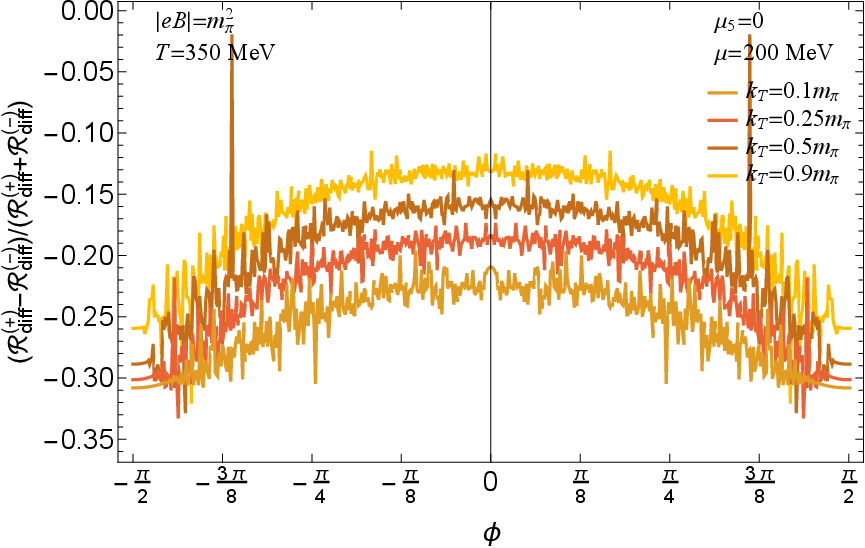}} 
 \hspace{0.05\textwidth}
{\includegraphics[width=0.44\textwidth]{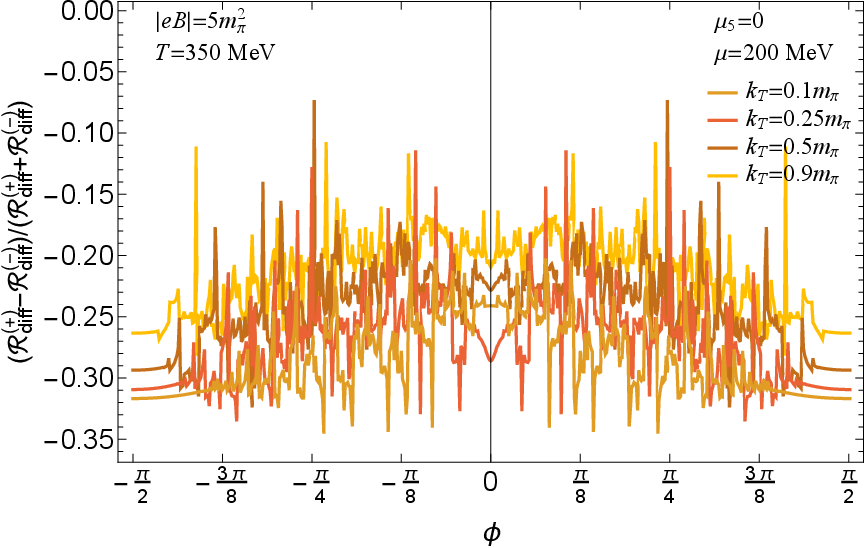}}
\caption{The degree of circular polarization as a function of the azimuthal angle for two magnetic fields,  $|eB| = m_\pi^2$ (left) and $|eB| = 5m_\pi^2$ (right), and two temperatures, $T = 200~\mbox{MeV}$ (top) and $T = 350~\mbox{MeV}$ (bottom).}
\label{fig:polarization-degree-mu}
\end{figure}

We observe, however, that the emission rates for left-handed and right-handed circular polarizations differ significantly, with the latter being a few times larger than the former. To quantify this effect, it is instructive to calculate the degree of circular polarization, defined by
\begin{eqnarray}
{\cal P}_{\rm circ}  (k_T,\phi, y)= \frac{{\cal R}^{(+)}_{\rm diff} (k_T,\phi, y) - {\cal R}^{(-)}_{\rm diff} (k_T,\phi, y)}{{\cal R}^{(+)}_{\rm diff} (k_T,\phi, y) + {\cal R}^{(-)}_{\rm diff} (k_T,\phi, y)}.
\label{P-circ}
\end{eqnarray}
The corresponding numerical results are presented in Fig.~\ref{fig:polarization-degree-mu}, which confirm a substantial degree of circular polarization in the photon emission. The sign of ${\cal P}_{\rm circ}  (k_T,\phi, y)$ is determined primarily by a nonzero electrical charge density in the plasma. Indeed, by separating the partial contributions from the up and down quarks, we find that the up quarks emit more left-handed photons (${\cal P}^{\rm (u)}_{\rm circ} <0$), while the down quarks emit more right-handed photons (${\cal P}^{\rm (d)}_{\rm circ} >0$). Since the up quarks have twice the electric charge of the down quarks, their emission rate is higher overall. Consequently, the net circular polarization from QGP is negative, ${\cal P}_{\rm circ} <0$. 

The effect of a nonzero chemical potential, resulting in a predominant emission of photons with one circular polarization over the other may resemble the underlying physics of helicons (whistlers). Recall that helicons are low-frequency electromagnetic excitations in strongly magnetized plasmas, driven by the Lorentz force \cite{Maxfield:1969aaa}. Their circular polarization is determined by the sign of the electric charge carriers  (e.g., electrons) with higher mobility, moving in an approximately static background of opposite charges (e.g., positive ions).

For the model parameters explored, the degree of circular polarization ranges from approximately $0.12$ to $0.5$.  It tends to be larger at stronger magnetic fields and lower transverse momenta (or, equivalently, lower photon energies). Comparing the top and bottom panels in Fig.~\ref{fig:polarization-degree-mu}, we also observe that the effect tends to diminish as the temperature rises.

\subsection{Nonzero chiral chemical potential $\mu_5$}
\label{subsec:mu5}

Let us now consider the case of a magnetized quark-gluon plasma (QGP) with a nonzero chiral chemical potential. Specifically, we choose $\mu_5 =100~\mbox{MeV}$ as a representative value, although the actual value is not crucial for identifying the qualitative effects of $\mu_5$. 

\begin{figure}[t]
\centering
{\includegraphics[width=0.44\textwidth]{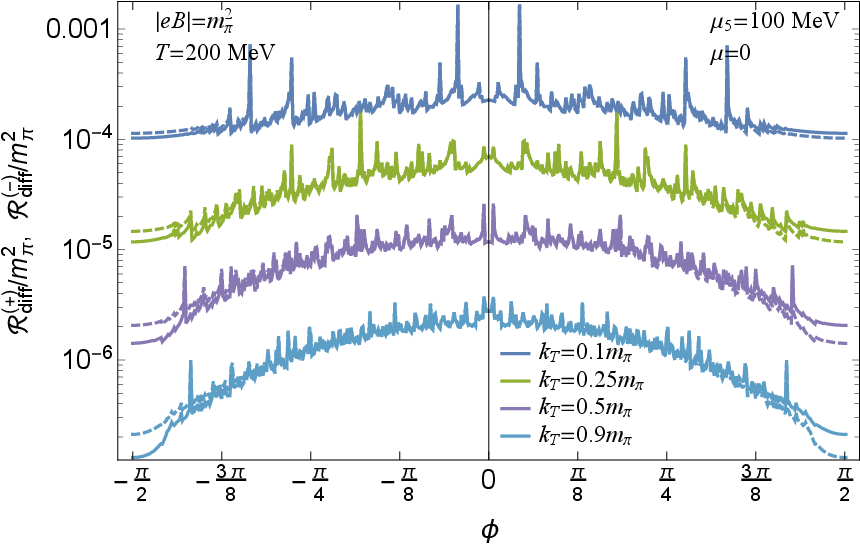}}
  \hspace{0.05\textwidth}
{\includegraphics[width=0.44\textwidth]{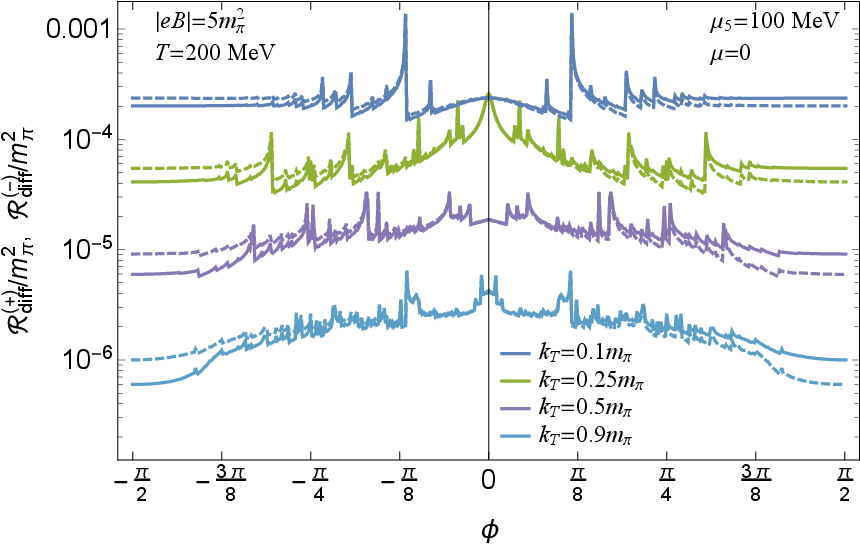}} 
\caption{The angular dependence of right-handed (solid lines) and left-handed (dashed lines) circularly polarized photon emission rates for two magnetic fields,  $|eB| = m_\pi^2$ (left) and $|eB| = 5m_\pi^2$ (right), and fixed temperature, $T = 200~\mbox{MeV}$.}
\label{fig:rates-mu5}
\end{figure}

Fig.~\ref{fig:rates-mu5} presents sample numerical results for the emission rates of left-handed and right-handed circularly polarized photons. The data is shown for a temperature of $T=200~\mbox{MeV}$ and for two different magnetic field strengths: $|eB| = m_\pi^2$ (left panel) and $|eB| = 5m_\pi^2$ (right panel). Although we do not show the rates at $T=350~\mbox{MeV}$, we have verified that they exhibit similar qualitative behavior, albeit with generally higher values.

\begin{figure}[t]
\centering
{\includegraphics[width=0.44\textwidth]{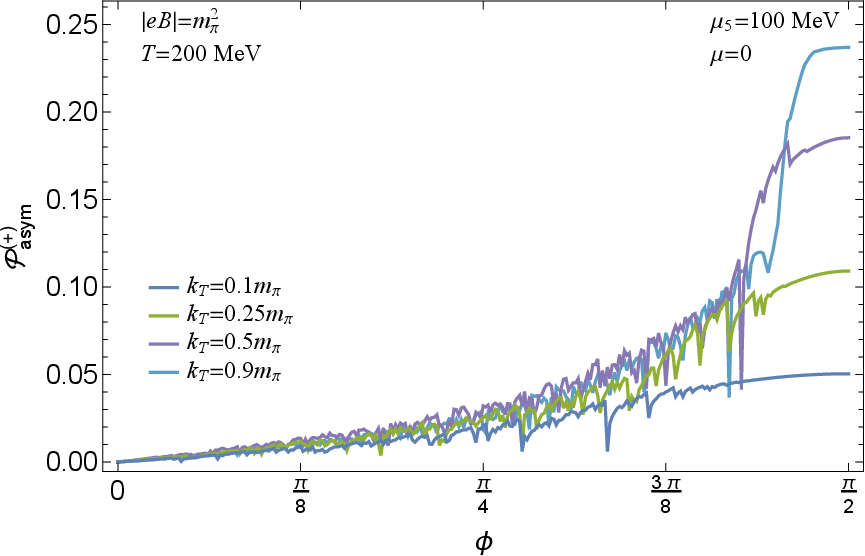}}
  \hspace{0.05\textwidth}
{\includegraphics[width=0.44\textwidth]{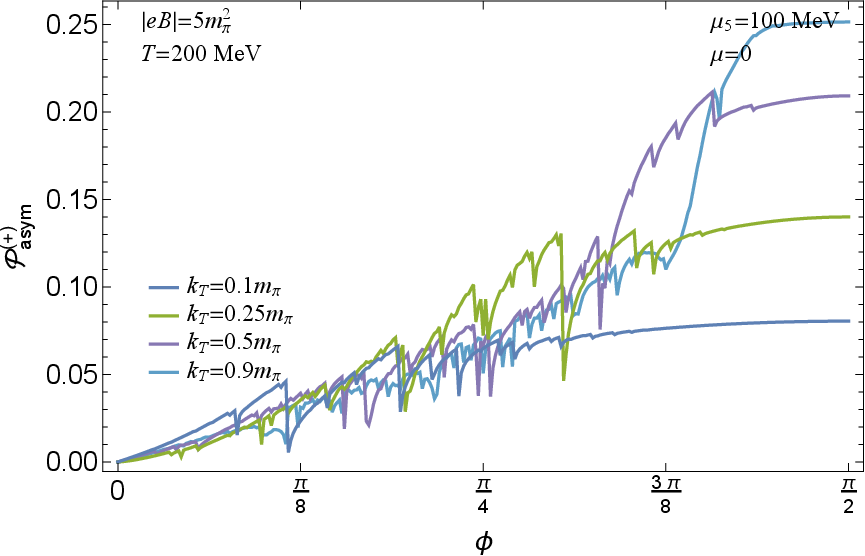}} \\[3mm]
{\includegraphics[width=0.44\textwidth]{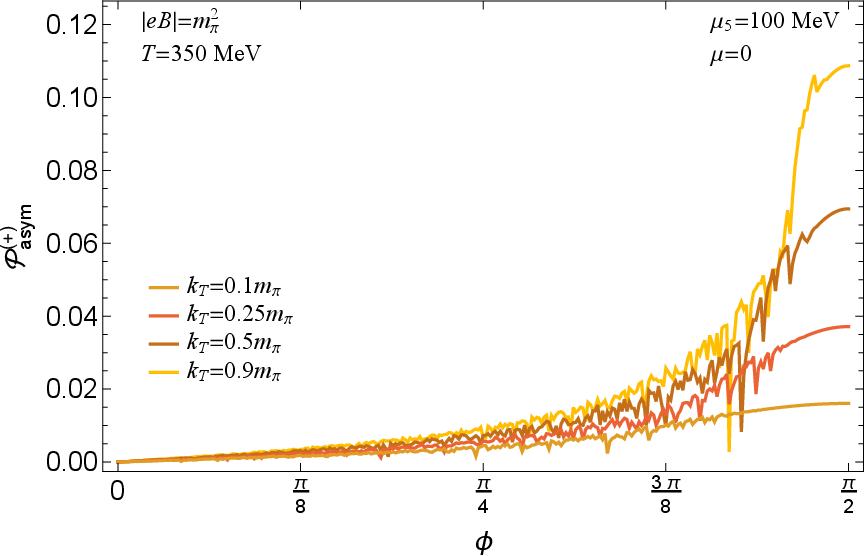}} 
  \hspace{0.05\textwidth}
{\includegraphics[width=0.44\textwidth]{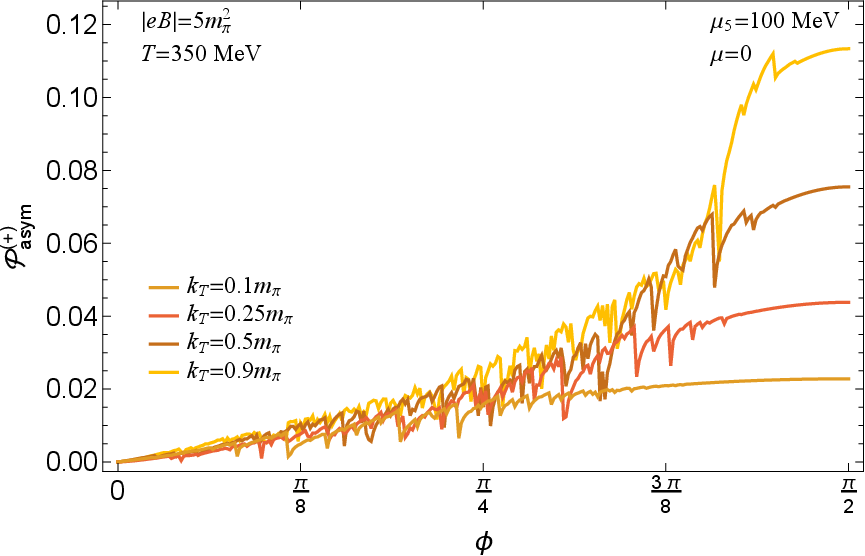}}
\caption{The emission asymmetry for the right-handed circularly polarized photons as a function of the azimuthal angle for two magnetic fields,  $|eB| = m_\pi^2$ (left) and $|eB| = 5m_\pi^2$ (right), and two temperatures, $T = 200~\mbox{MeV}$ (top) and $T = 350~\mbox{MeV}$ (bottom).}
\label{fig:asym-mu5}
\end{figure}

The emission rates in Fig.~\ref{fig:rates-mu5} possess an interesting property, namely they are asymmetric with respect to reflection in the transverse plane (i.e., $\phi\to -\phi$). To visualize this effect, we construct the following two observables:
\begin{eqnarray}
{\cal P}^{(+)}_{\rm asym} (k_T,\phi, y)&=&\frac{{\cal R}^{(+)}_{\rm diff} (k_T,\phi, y) - {\cal R}^{(+)}_{\rm diff} (k_T,-\phi, y)}{{\cal R}^{(+)}_{\rm diff} (k_T,\phi, y) + {\cal R}^{(+)}_{\rm diff} (k_T,-\phi, y)},
\label{P-plus-asym}
\\
{\cal P}^{(-)}_{\rm asym}  (k_T,\phi, y)&=&\frac{{\cal R}^{(-)}_{\rm diff} (k_T,\phi, y) - {\cal R}^{(-)}_{\rm diff} (k_T,-\phi, y)}{{\cal R}^{(-)}_{\rm diff} (k_T,\phi, y) + {\cal R}^{(-)}_{\rm diff} (k_T,-\phi, y)},
\label{P-minus-asym}
\end{eqnarray}
to measure the degree of emission asymmetry for the left-handed and the right-handed circularly polarized photons, respectively. It is sufficient to   define them only for positive azimuthal angles in the range $0\leq \phi \leq \pi/2$. The corresponding numerical results for the emission asymmetry ${\cal P}^{(+)}_{\rm asym} (k_T,\phi, y)$ are presented in Fig.~\ref{fig:asym-mu5} for two magnetic fields, $|eB| = m_\pi^2$ (left) and $|eB| = 5m_\pi^2$ (right), and two temperatures, $T = 200~\mbox{MeV}$ (top) and $T = 350~\mbox{MeV}$ (bottom). We do not show any numerical data for ${\cal P}^{(-)}_{\rm asym} (k_T,\phi, y)$ because it is not truly independent. Indeed, ${\cal P}^{(-)}_{\rm asym}  (k_T,\phi, y) = - {\cal P}^{(+)}_{\rm asym}  (k_T,\phi, y)$. As we will see later, this relation will not remain valid at nonzero $\mu$. 

Fig~\ref{fig:asym-mu5} shows that for a positive $\mu_5$, the emission asymmetry is positive for photons with the right-handed circular polarization. Since ${\cal P}^{(-)}_{\rm asym}  (k_T,\phi, y) = - {\cal P}^{(+)}_{\rm asym}  (k_T,\phi, y)$, the asymmetry is negative for photons with left-handed circular polarization. This implies that the emission rate for right-handed polarized photons is higher in the direction along the magnetic field, while for left-handed polarized photons, it is higher in the direction opposite to the magnetic field. Additionally, we find that the degree of asymmetry is most pronounced at $\phi\simeq \pi/2$ and that its magnitude increases with increasing transverse momentum. We also observe that the asymmetry grows with increasing magnetic field strength but diminishes with rising temperature.

\subsection{Nonzero quark-number and chiral chemical potentials $\mu$ and $\mu_5$}
\label{subsec:mu-mu5}

After considering the effects of $\mu$ and $\mu_5$ separately in the previous two subsections, let us now study their combined effect on polarized photon emission. As one might expect, the emission is characterized by a combined set of the same qualitative features, namely (i) the overall degree of circular polarization, indicating that the emission of one circular polarization dominates over the other and (ii) the degree of asymmetry for each circular polarization with respect $\phi\to -\phi$, measuring how asymmetric is the emission rate with respect to reflection in the transverse plane. 

To quantify the effects in the case of nonzero $\mu$ and $\mu_5$, we use the same set of observables that we introduced earlier, namely ${\cal P}_{\rm circ}  (k_T,\phi, y)$ in Eq.~(\ref{P-circ}), ${\cal P}^{(+)}_{\rm asym} (k_T,\phi, y)$ in Eq.~(\ref{P-plus-asym}), and ${\cal P}^{(-)}_{\rm asym}  (k_T,\phi, y)$ in  Eq.~(\ref{P-minus-asym}). 

\begin{figure}[t]
\centering
{\includegraphics[width=0.44\textwidth]{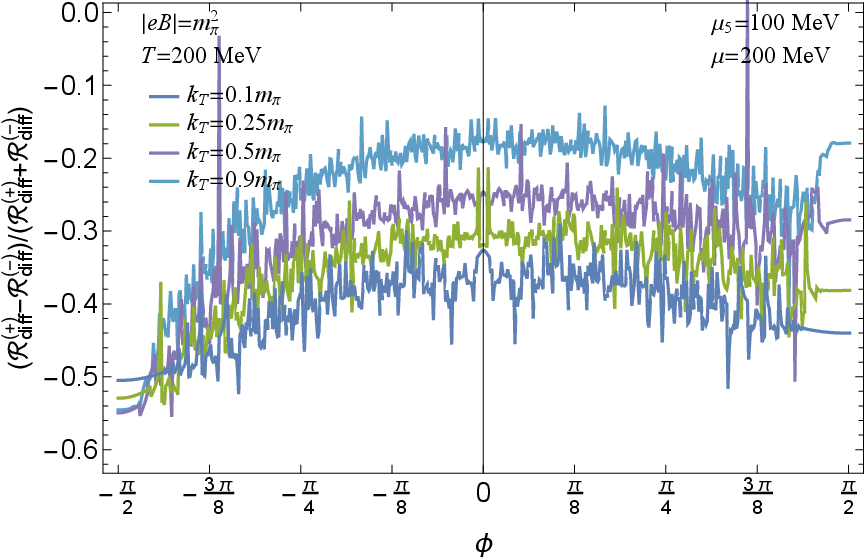}}
  \hspace{0.05\textwidth}
{\includegraphics[width=0.44\textwidth]{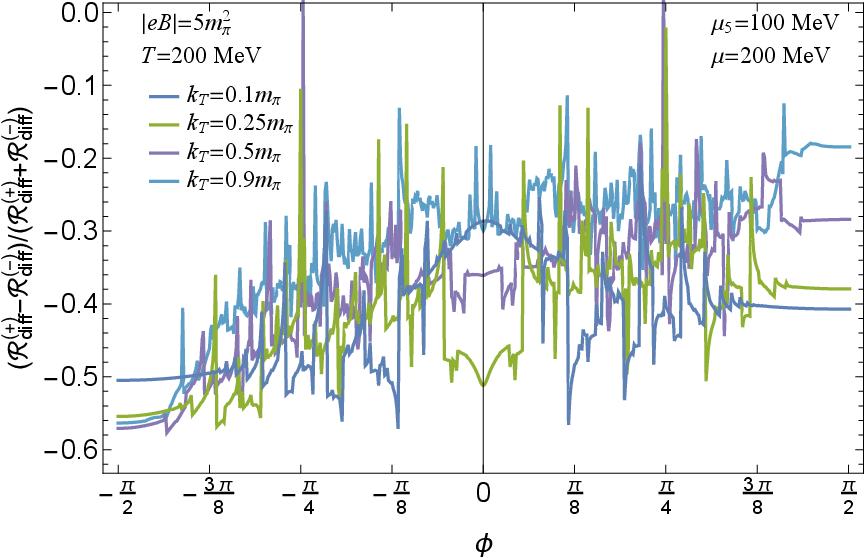}} \\[3mm]
{\includegraphics[width=0.44\textwidth]{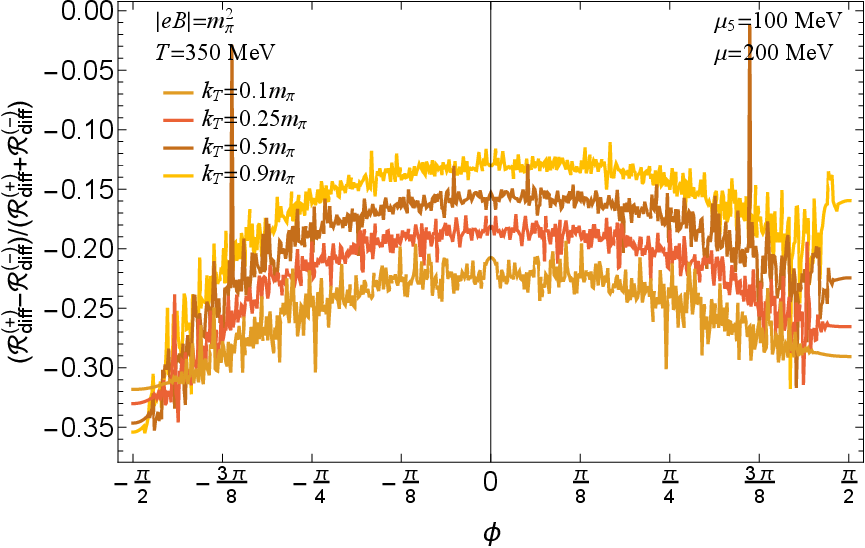}}
\hspace{0.05\textwidth}
{\includegraphics[width=0.44\textwidth]{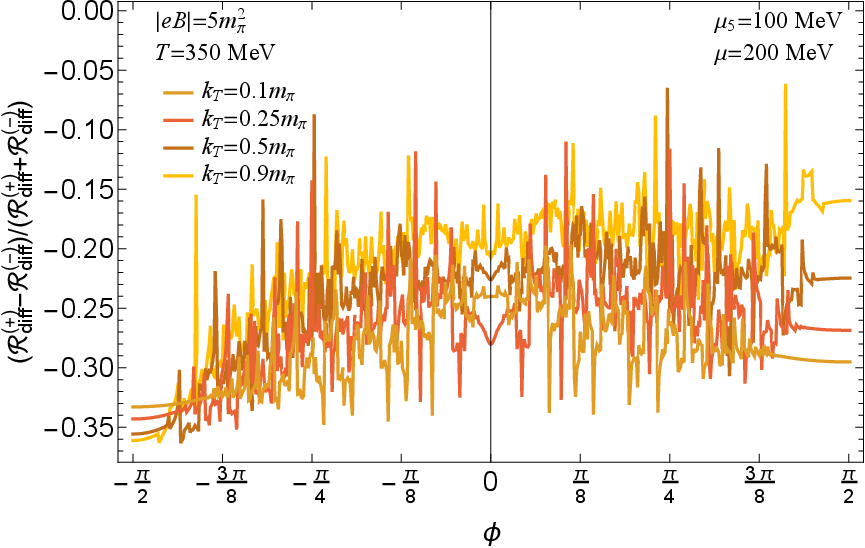}}
\caption{The degree of circular polarization as a function of the azimuthal angle for two magnetic fields, $|eB| = m_\pi^2$ (left) and $|eB| = 5m_\pi^2$ (right), and two temperatures, $T = 200~\mbox{MeV}$ (top) and $T = 350~\mbox{MeV}$ (bottom).}
\label{fig:dop-mu-mu5}
\end{figure}

 The numerical results for the overall degree of circular polarization in photon emission are presented in Fig.~\ref{fig:dop-mu-mu5}. The magnitude of the effect is comparable to that discussed in Sec.~\ref{subsec:mu}, and the qualitative dependence on the strength of the magnetic field, temperature, and transverse momentum is similar. However, there is a significant difference too. The angular dependence is not symmetric under the reflection $\phi\to -\phi$. The degree of circular polarization is larger in magnitude in the direction opposite to the magnetic field. Of course, this is expected since the dominant left-handed emission is asymmetric. 

\begin{figure}[t]
\centering
{\includegraphics[width=0.44\textwidth]{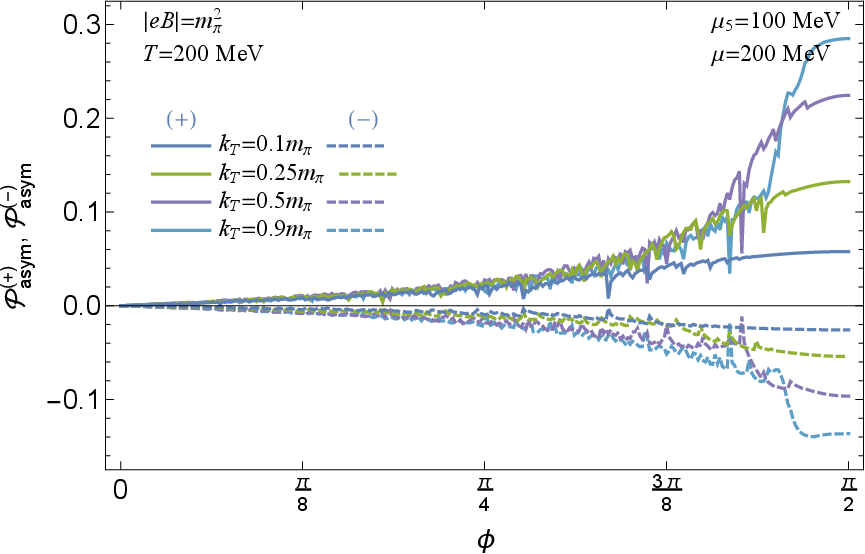}} 
  \hspace{0.05\textwidth}
{\includegraphics[width=0.44\textwidth]{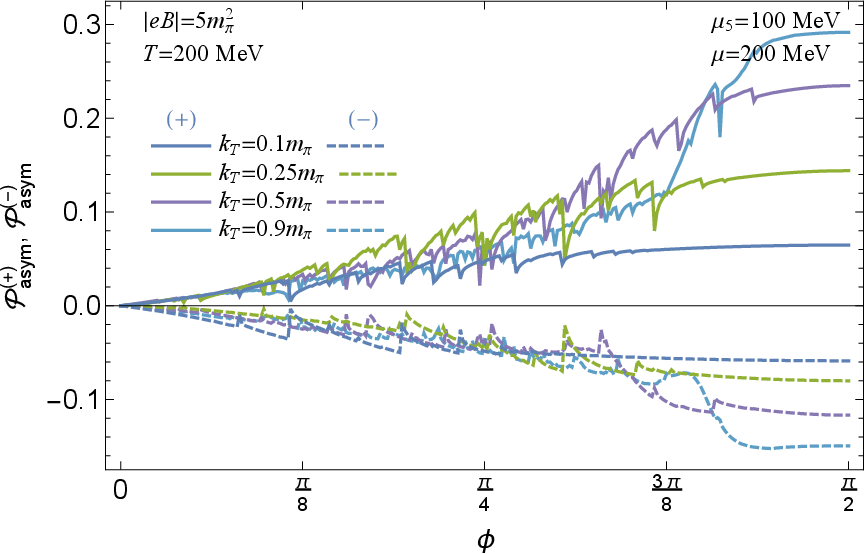}}\\[3mm]
{\includegraphics[width=0.44\textwidth]{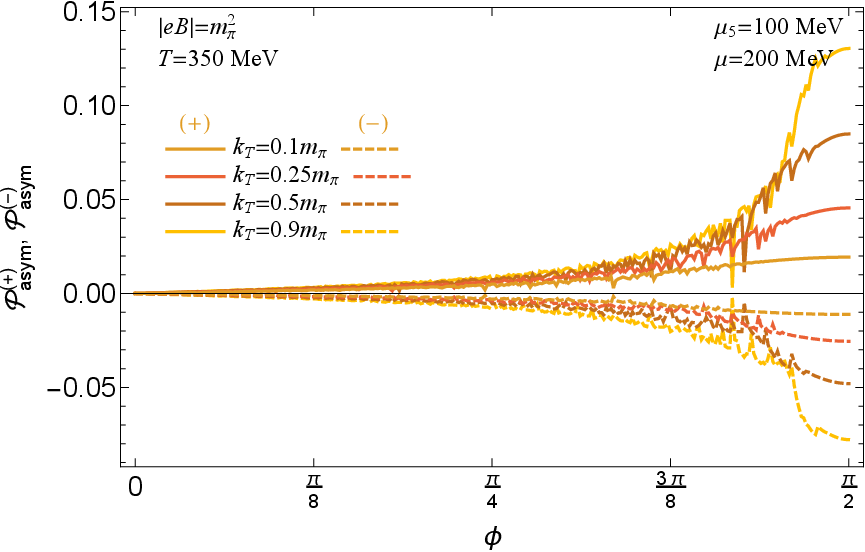}} 
  \hspace{0.05\textwidth}
{\includegraphics[width=0.44\textwidth]{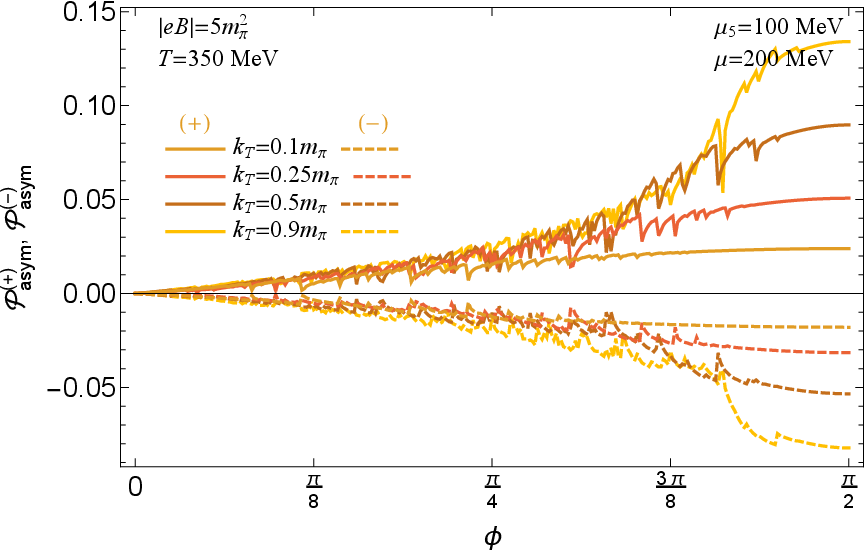}} 
\caption{The emission asymmetry for the right-handed (solid lines) and left-handed (dashed lines) circularly polarized photons as a function of the azimuthal angular for two magnetic fields,  $|eB| = m_\pi^2$ (left) and $|eB| = 5m_\pi^2$ (right), and two temperatures, $T = 200~\mbox{MeV}$ (top) and $T = 350~\mbox{MeV}$ (bottom).}
\label{fig:asym-mu-mu5}
\end{figure}

The degrees of asymmetry for the emission of photons of both circular polarizations are shown in Fig.~\ref{fig:asym-mu-mu5} for two magnetic fields,  $|eB| = m_\pi^2$ (left panels) and $|eB| = 5m_\pi^2$ (right panels), and two temperatures, $T = 200~\mbox{MeV}$ (top panels) and $T = 350~\mbox{MeV}$ (bottom panels). For both temperatures, the degree of asymmetry for the right-handed emission is larger than for the left-handed emission. The largest asymmetry is seen around $\phi\simeq \pi/2$, with the magnitudes lying in the range from about $0.12$ (right-handed) to $0.3$ (left-handed) at $T=200~\mbox{MeV}$ and from about $0.08$ (right-handed) to $0.13$ (left-handed) at $T=350~\mbox{MeV}$. 

\section{Discussion and Summary}
\label{Summary}

In this study, we investigated the effect of nonzero quark-number (electric charge) and chiral chemical potentials on the polarization of photon emission from a  hot, strongly magnetized QGP. Under the assumption of sufficiently strong background magnetic fields and high temperatures, photon emission is dominated by leading-order one-to-two and two-to-one processes, as shown in Fig.~\ref{fig:processes}, while contributions from gluon-mediated two-to-two processes are neglected as subleading.

This work extends previous studies in Refs.~\cite{Wang:2020dsr,Wang:2021ebh,Wang:2021eud} that employed the same approximation but focused exclusively on unpolarized photon emission from QGP with vanishing chiral chemical potential. In contrast, here we demonstrate that the composition of circularly polarized photon emission undergoes qualitative changes when nonzero chiral charge density is introduced. More interestingly, the emission of left-handed and right-handed polarized photons can be partially separated (or ``distilled") into the regions  above and below the reaction plane.

A nonzero quark-number (electrical charge) density produces an overall dominance of one circular polarization over the other. For example, when the QGP has a positive quark-number (electrical charge) chemical potential ($\mu> 0$), the emission rate of left-handed circular polarization photons dominates over the rate of right-handed circular polarization photons. In effect, such polarized emission can be viewed as a photon equivalent of the Hall effect in a plasma with a nonzero charge density. 

It should be mentioned that the circularly polarized photon emission rates are symmetric with respect to the reflection in the transverse (reaction) plane when the electrical charge density is nonzero ($\mu\neq 0$), assuming the chiral charge density vanishes ($\mu_5= 0$). However, this changes at nonzero chiral charge density ($\mu_5\neq 0$). When $\mu_5> 0$, we show that the emission rate of right-handed (left-handed) circularly polarized photons is higher in the upper (lower) hemisphere. (Here the two hemispheres are defined by the magnetic field direction.) The roles of the two circular polarizations interchange when the sign of $\mu_5$ changes. 

As expected, the combined effect of nonzero electrical and chiral charge densities in the QGP is given by a superposition of their individual effects. Generally, a nonzero electrical charge leads to an overall dominance of the photon emission with one circular polarization over the other. The additional chiral charge density produces spatial asymmetries (with respect to reflection in the transverse plane) in photon emission for each circular polarization. Moreover, while the rate for one circular polarization is higher in the upper hemisphere, the rate for the other circular polarization is higher in the lower hemisphere. 

Our findings suggest that the composition of circularly polarized photon emission and its asymmetry with respect to reflection in the transverse plane can be used as unambiguous observable signatures of nonzero electrical and chiral charge densities in a strongly magnetized plasma. 

\acknowledgments{We would like to thank Kirill Tuchin for valuable comments regarding the early draft of the paper. 
The work of X.~W. was supported by Anhui University of Science and Technology under Grant No.~YJ20240001. 
The work of I.~A.~S. was supported in part by the U.S. National Science Foundation under Grant No.~PHY-2209470.}

\appendix

\section{Matsubara sums}
\label{app:MatsubaraSums}

In the calculation of the polarization tensor in the main text, we perform Matsubara sums by using the following general result:  
\begin{eqnarray}
&& T\sum_{k=-\infty}^{\infty} 
\frac{(i\omega_k+\mu+s\mu_5) (i\omega_k-i\Omega_m+\mu+s\mu_5) X + (i\omega_k+\mu+s\mu_5) Y_1 +(i\omega_k-i\Omega_m+\mu+s\mu_5) Y_2 +Z}
{\left[(i\omega_k+\mu+s\mu_5)^2-a^2\right]\left[(i\omega_k-i\Omega_m+\mu+s\mu_5)^2-b^2\right]} \nonumber\\
&&= \sum_{\lambda=\pm 1} \sum_{\eta=\pm 1}
\frac{\left[ n_F(a+\eta \mu+\eta s\mu_5)-n_F(\lambda b+\eta \mu+\eta s\mu_5)  \right]}{4 \lambda ab\left(a-\lambda b+ \eta  i\Omega_m\right)}
\left[\lambda ab X-\eta\left( a Y_1+\lambda bY_2 \right)+Z\right],
\label{Matsubara-sum}
\end{eqnarray}
where coefficient $X$, $Y_1$, $Y_2$, and $Z$ are arbitrary functions of momenta. In the calculation of the polarization tensor, parameters $a$ and $b$ are replaced with the Landau level energies, $E_{n,p_z,f}=\sqrt{p_z^2+2n|e_f B|}$ and $E_{n^{\prime},p_z-k_z,f}=\sqrt{(p_z-k_z)^2+2n^{\prime}|e_fB|}$.

Note that the expression in the parenthesis of the final result in Eq.~(\ref{Matsubara-sum}) can be formally obtained from 
the numerator of the original expression by making the following replacements: 
$(i\omega_k+s\mu_5)(i\omega_k-i\Omega_m+s\mu_5)\to \lambda ab$,
$(i\omega_k+s\mu_5) \to  -\eta a$,
and $(i\omega_k-i\Omega_m+s\mu_5)\to -\eta\lambda b$.

\section{Calculation of Dirac traces}
\label{app:Traces}

In the derivation of the photon polarization function in Sec.~\ref{sec:Polarization}, one has to calculate four different types of Dirac traces, i.e.,
\begin{eqnarray}
\label{trt1}
T_{1,f}^{\mu\nu}(s_1,s_2) &=&
\tr \left\{ \gamma^\mu \left(\bar{p}_\parallel \gamma_\parallel -\mu_5 \gamma^0\gamma^5\right)\left( \mathcal{P}_{+}^fL_n +\mathcal{P}_{-}^fL_{n-1} \right)\mathcal{P}_5^{s_1}\gamma^\nu \left((\bar{p}_\parallel-k_\parallel )\gamma_\parallel  -\mu_5 \gamma^0\gamma^5 \right)\left( \mathcal{P}_{+}^fL_{n^\prime} +\mathcal{P}_{-}^fL_{n^\prime-1} \right)\mathcal{P}_5^{s_2}\right\}, \nonumber \\
\label{trt2} \\
T_{2,f}^{\mu\nu} (s_1,s_2)&=& \frac{i}{\ell_f^2}
\tr \left\{ \gamma^\mu \left(\bar{p}_\parallel\gamma_\parallel -\mu_5 \gamma^0\gamma^5\right)\left( \mathcal{P}_{+}^fL_n +\mathcal{P}_{-}^fL_{n-1} \right)\mathcal{P}_5^{s_1} \gamma^\nu  (\mathbf{r}_{\perp}\cdot\bm{\gamma}_{\perp}) L_{n^\prime-1}^{1}\mathcal{P}_5^{s_2}\right\}, \\
\label{trt3}
T_{3,f}^{\mu\nu}(s_1,s_2) &=& -\frac{i}{\ell_f^2}
\tr \left\{ \gamma^\mu (\mathbf{r}_{\perp}\cdot\bm{\gamma}_{\perp}) L_{n-1}^{1} \mathcal{P}_5^{s_1}
\gamma^\nu  \left[(\bar{p}_\parallel-k_\parallel )\gamma_\parallel  -\mu_5 \gamma^0\gamma^5\right]
\left( \mathcal{P}_{+}^fL_{n^\prime} +\mathcal{P}_{-}^fL_{n^\prime-1} \right)\mathcal{P}_5^{s_2} \right\},\\
\label{trt4}
T_{4,f}^{\mu\nu}(s_1,s_2) &=&\frac{1}{\ell_f^4}
\tr \left\{ \gamma^\mu (\mathbf{r}_{\perp}\cdot\bm{\gamma}_{\perp}) L_{n-1}^{1}   \mathcal{P}_5^{s_1}\gamma^\nu  (\mathbf{r}_{\perp}\cdot\bm{\gamma}_{\perp}) L_{n^\prime-1}^{1}\mathcal{P}_5^{s_2} \right\}.
\end{eqnarray}
Here we used the shorthand notation $\bar{p}_\parallel \gamma_\parallel =\bar{p}_0\gamma^0-p^3\gamma^3$ and $\bar{p}_0=p_0+\mu$. Also, for brevity of presentation, we omitted the explicit dependence of all functions on their arguments, i.e.,
$L_{n}^{\alpha} \equiv L_{n}^{\alpha}\left(\zeta\right)$, where $\zeta=r_\perp^2/(2\ell_f^2)$.

For the purposes of this study, we must separate the results of traces in Eqs.~(\ref{trt1}) -- (\ref{trt4}) into individual polarization contributions. By using the definition in Eq.~(\ref{decompose}), we introduce the following polarization projections:
\begin{eqnarray}
\mathcal{T}_{0}^{f}(s_1,s_2)&=&\epsilon_{0}^\mu  \epsilon_{0}^{\nu*}\sum_{i=1}^4T_{i,f}^{\mu\nu}(s_1,s_2), \\
\mathcal{T}_{-}^{f}(s_1,s_2)&=&-\epsilon_{-}^\mu  \epsilon_{-}^{\nu*}\sum_{i=1}^4T_{i,f}^{\mu\nu}(s_1,s_2),\\
\mathcal{T}_{+}^{f}(s_1,s_2)&=&-\epsilon_{+}^\mu  \epsilon_{+}^{\nu*}\sum_{i=1}^4T_{i,f}^{\mu\nu}(s_1,s_2),\\
\mathcal{T}_{\parallel}^{f}(s_1,s_2)&=&\epsilon_{\parallel}^\mu  \epsilon_{\parallel}^{\nu*}\sum_{i=1}^4T_{i,f}^{\mu\nu}(s_1,s_2).
\end{eqnarray}
For each function, the result is nonzero only when $s_1=s_2$. The corresponding nonvanishing results read
\begin{eqnarray}
\mathcal{T}_{0}^{f}(s,s)&=& \left[ ( \bar{p}_0- s \mu_5 -p_z s_\perp s ) ( \bar{p}_0 - k_0-s\mu_5 )
+(p_z-k_z) \left( p_z - s_\perp s \bar{p}_0 + s_\perp \mu_5  \right) \right]  L_{n}L_{n^\prime}
\nonumber\\
& +& \left[ ( \bar{p}_0-s\mu_5 +p_z s_\perp s  ) ( \bar{p}_0 - k_0-s\mu_5 )
+(p_z-k_z)  (p_z+ s_\perp s  \bar{p}_0- s_\perp s \mu_5  ) \right]  L_{n-1}L_{n^\prime-1}  
\nonumber\\
& +& \frac{2 r_\perp^2}{\ell_f^4}L_{n-1}^1L_{n^\prime-1}^1, \\
\mathcal{T}_{-}^{f}(s,s)
&=&(s \bar{p}_0+ p_z-\mu_5) ( s k_0- s \bar{p}_0-k_z+p_z+\mu_5) 
\left[ L_{n-1}L_{n^\prime}(1+s_\perp)+ L_{n}L_{n^\prime-1}(1-s_\perp) \right], \\
\mathcal{T}_{+}^{f}(s,s)
&=&(s \bar{p}_0-p_z-\mu_5) ( s k_0- s \bar{p}_0+k_z-p_z+\mu_5) 
\left[ L_{n}L_{n^\prime-1}(1+s_\perp)+ L_{n-1}L_{n^\prime}(1-s_\perp)\right], \\
\mathcal{T}_{\parallel}^{f}(s,s)&=&-\mathcal{T}_{0}^{f}(s,s)+\frac{4 L_{n-1}^1L_{n^\prime-1}^1r_\perp^2}{\ell_f^4}.
\end{eqnarray}

\section{Integration over transverse spatial coordinates ($\mathbf{r}_\perp$)}
\label{app:Integral-R-perp}

After substituting the results for the Dirac traces into the definition of the polarization tensor in Eq.~(\ref{Pi_Omega_k-alt}), we have to integrate the resulting expressions over the transverse spatial coordinates. There corresponding integrals for the four different polarization components are given by:
\begin{eqnarray}
\mathcal{J}^{f}_{0}(s,s)&=& \int   d^2 \mathbf{r}_\perp e^{-i \mathbf{r}_\perp\cdot \mathbf{k}_\perp} e^{-\mathbf{r}_\perp^2/(2\ell_f^2)} \mathcal{T}_{0}^{f}(s,s)\nonumber\\
&=&2\pi\ell_f^2 \left[( \bar{p}_0-s \mu_5 -p_z s_\perp s ) ( \bar{p}_0 - k_0-s\mu_5 )
+(p_z-k_z) \left( p_z - s_\perp s \bar{p}_0 + s_\perp \mu_5  \right) \right]  \, \mathcal{I}_{0,f}^{n,n^\prime}(\xi)
\nonumber\\
&+& 2\pi \ell_f^2 \left[ ( \bar{p}_0-s\mu_5 +p_z s_\perp s  ) ( \bar{p}_0 - k_0-s\mu_5 )
+(p_z-k_z)  (p_z+ s_\perp s  \bar{p}_0- s_\perp s \mu_5  ) \right]  \, \mathcal{I}_{0,f}^{n-1,n^\prime-1}(\xi)
\nonumber\\
&+& 8\pi \, \mathcal{I}_{2,f}^{n-1,n^\prime-1}(\xi),\\
\mathcal{J}^{f}_{-}(s,s)&=& \int   d^2 \mathbf{r}_\perp e^{-i \mathbf{r}_\perp\cdot \mathbf{k}_\perp} e^{-\mathbf{r}_\perp^2/(2\ell_f^2)} \mathcal{T}_{+}^{f}(s,s)\nonumber\\
&=&2\pi\ell_f^2 (s \bar{p}_0+ p_z-\mu_5) ( s k_0- s \bar{p}_0-k_z+p_z+\mu_5) 
\left[(1+s_\perp)\mathcal{I}_{0,f}^{n-1,n^\prime}(\xi)+(1-s_\perp)\mathcal{I}_{0,f}^{n,n^\prime-1}(\xi)\right],\\
\mathcal{J}^{f}_{+}(s,s)&=& \int   d^2 \mathbf{r}_\perp e^{-i \mathbf{r}_\perp\cdot \mathbf{k}_\perp} e^{-\mathbf{r}_\perp^2/(2\ell_f^2)} \mathcal{T}_{-}^{f}(s,s)\nonumber\\
&=&2\pi \ell_f^2  (s \bar{p}_0-p_z-\mu_5) ( s k_0- s \bar{p}_0+k_z-p_z+\mu_5)  
\left[(1+s_\perp)\mathcal{I}_{0,f}^{n,n^\prime-1}(\xi)+(1-s_\perp)\mathcal{I}_{0,f}^{n-1,n^\prime}(\xi)\right],\\
\mathcal{J}^{f}_{\parallel}(s,s)&=& \int   d^2 \mathbf{r}_\perp e^{-i \mathbf{r}_\perp\cdot \mathbf{k}_\perp} e^{-\mathbf{r}_\perp^2/(2\ell_f^2)} \mathcal{T}_{\parallel}^{f}(s,s)=8\pi \, \mathcal{I}_{2,f}^{n-1,n^\prime-1}(\xi)-\mathcal{J}^{f}_{0}(s,s) , 
\end{eqnarray}
where $\xi = k_{\perp}^2\ell_f^{2}/2$ and functions
\begin{eqnarray}
\label{I-0} 
\mathcal{I}_{0,f}^{n,n^{\prime}}(\xi)&=& \frac{(n^\prime)!}{n!} e^{-\xi}  \xi^{n-n^\prime} \left(L_{n^\prime}^{n-n^\prime}\left(\xi\right)\right)^2 
= \frac{n!}{(n^\prime)!}e^{-\xi} \xi^{n^\prime-n} \left(L_{n}^{n^\prime-n}\left(\xi\right)\right)^2, \\
\label{I-2} 
\mathcal{I}_{2,f}^{n,n^{\prime}}(\xi)&=& \frac{n+n^{\prime}+2}{2}\left[\mathcal{I}_{0,f}^{n,n^{\prime}}(\xi) +\mathcal{I}_{0,f}^{n+1,n^{\prime}+1}(\xi) \right]  -\frac{\xi}{2}\left[\mathcal{I}_{0,f}^{n+1,n^{\prime}}(\xi) +\mathcal{I}_{0,f}^{n,n^{\prime}+1}(\xi) \right] 
\end{eqnarray}
are the same as those introduced in Ref.~\cite{Wang:2021ebh}.

\section{Imaginary part of $\Pi^{\mu\nu}(\Omega;\mathbf{k})$}
\label{app:ImaginaryPart}

By making use of the results in Appendices~\ref{app:Traces} and \ref{app:Integral-R-perp}, we derive the following expressions for the polarization projections of the imaginary part of $\Pi^{\mu\nu}(i\Omega_m;\mathbf{k})$:
\begin{equation}
\Pi^{(i)}(i\Omega_m;\mathbf{k})=  - \sum_{f=u,d}\frac{N_c \alpha_f T}{\pi \ell_f^4} \sum_{n,n^\prime = 0}^{\infty} \sum_{k=-\infty}^{\infty} \sum_{s=\pm 1}\int \frac{dp_z}{2\pi}\frac{\mathcal{J}_{(i)}^{f}(-s,-s)}{[(i\omega_k+\mu+s\mu_5)^2-E_{n,p_z,f}^2][(i\omega_k+\mu-i\Omega_m+s\mu_5)^2-E_{n^\prime,p_z-k_z,f}^2]  }. 
\end{equation}
By using a general result for the Matsubara sum in Appendix~\ref{app:MatsubaraSums}, we derive
\begin{equation}
\Pi^{(i)}(i\Omega_m;\mathbf{k})=-\sum_{f=u,d} \frac{N_c \alpha_f }{\pi \ell_f^4} \sum_{n,n^\prime = 0}^{\infty} \sum_{s,\eta,\lambda=\pm 1}\int \frac{dp_z}{2\pi}\frac{[n_F(E_{n,p_z,f}+\eta \mu+ \eta s \mu_5)-n_F(\lambda E_{n^\prime,p_z-k_z,f}+ \eta \mu+\eta s \mu_5)]\mathcal{F}^{(i)}_{s,f}(\xi, i\Omega_m)}{4\lambda E_{n,p_z,f}E_{n^\prime,p_z-k_z,f}\left[E_{n,p_z,f}-\lambda E_{n^\prime,p_z-k_z,f}+i \eta \Omega_m \right]},
\end{equation}
where
\begin{eqnarray}
\mathcal{F}^{(0)}_{s,f}(\xi, i\Omega_m) &=&2\pi\ell_f^2 \left(p_z - s_{\perp} s \eta E_{n,p_z,f}\right) 
\left[ p_z - k_z -s_\perp s \eta \lambda E_{n^\prime,p_z-k_z,f}\right] \mathcal{I}_{0,f}^{n,n^\prime}(\xi)\nonumber\\
&+&2\pi\ell_f^2  \left(p_z + s_{\perp} s \eta E_{n,p_z,f}\right) 
\left[p_z - k_z +s_\perp s \eta \lambda E_{n^\prime,p_z-k_z,f}\right]
 \mathcal{I}_{0,f}^{n-1,n^\prime-1}(\xi)
 + 4\pi\mathcal{I}_{2,f}^{n-1,n^{\prime }-1}(\xi), 
\label{F-0}\\
\mathcal{F}^{(-)}_{s,f}(\xi, i\Omega_m) &=& 2\pi\ell_f^2 \left[p_z (p_z-k_z)- k_z s \eta E_{n,p_z,f} -p_z s i \Omega_m-\lambda E_{n,p_z,f}E_{n^\prime,p_z-k_z,f}\right]\nonumber\\
&\times&\left[(1+s_\perp)\mathcal{I}_{0,f}^{n-1,n^\prime}(\xi)+(1-s_\perp)\mathcal{I}_{0,f}^{n,n^\prime-1}(\xi)\right],
\label{F-minus}\\ 
\mathcal{F}^{(+)}_{s,f}(\xi, i\Omega_m) &=& 2\pi\ell_f^2 \left[p_z(p_z-k_z) +k_z s \eta E_{n,p_z,f}+ p_z s i\Omega_m - \lambda E_{n,p_z,f}E_{n^\prime,p_z-k_z,f}\right]\nonumber\\
&\times&\left[(1+s_\perp)\mathcal{I}_{0,f}^{n,n^\prime-1}(\xi)+(1-s_\perp)\mathcal{I}_{0,f}^{n-1,n^\prime}(\xi)\right], 
\label{F-plus} \\
\mathcal{F}^{(\parallel)}_{s,f}(\xi, i\Omega_m) &=&-\mathcal{F}^{(0)}_{s,f}(\xi, i\Omega_m) +8\pi\mathcal{I}_{2,f}^{n-1,n^{\prime }-1}(\xi). 
\label{F-par}
\end{eqnarray}
After replacing $i\Omega_m \to \Omega+i\epsilon$ and using the Sokhotski formula, we extract the following imaginary part of the polarization functions:
\begin{eqnarray}
\mbox{Im}\Pi^{(i)} (\Omega;\mathbf{k}) &=&\sum_{f=u,d}\frac{N_c \alpha_f }{\ell_f^4} \sum_{n,n^\prime = 0}^{\infty} \sum_{s,\eta,\lambda=\pm 1}\int \frac{dp_z}{2\pi}\frac{[n_F(E_{n,p_z,f}+\eta \mu+\eta s \mu_5)-n_F(\lambda E_{n^\prime,p_z-k_z,f}+ \eta \mu+\eta s \mu_5)]}{4\lambda E_{n,p_z,f}E_{n^\prime,p_z-k_z,f}}\nonumber\\
&\times&  \mathcal{F}_{s,f}^{(i)}(\xi,\Omega) \, \delta\left(E_{n,p_z,f}-\lambda E_{n^\prime,p_z-k_z,f}+ \eta \Omega \right).
\label{eqE7}
\end{eqnarray}
The solutions of the energy conservation equation $E_{n,p_z,f}-\lambda E_{n^{\prime},p_z-k_z,f}+ \eta\Omega=0$ are given by the following explicit expressions~\cite{Wang:2020dsr,Wang:2021ebh,Wang:2021eud}:
\begin{eqnarray}
p_{z,f}^{(\pm)}&=& \frac{k_z}{2}\left[1+ \frac{2(n-n^{\prime})|e_fB|}{\Omega^2-k_z^2} 
\pm \frac{\Omega}{|k_z|}  \sqrt{\left(1-\frac{(k_{-}^{f})^2}{\Omega^2-k_z^2} \right)
\left( 1-\frac{(k_{+}^{f})^2}{\Omega^2-k_z^2}\right)} \right].
\label{pz-solution}
\end{eqnarray}
The corresponding fermions energies, satisfying the energy conservation relation, are 
\begin{subequations}
\begin{equation}
 \left. E_{n,p_z,f}\right|_{p_z = p_{z,f}^{(\pm)} } = - \frac{\eta \Omega}{2} \left[
1+\frac{2(n-n^{\prime})|e_fB|}{\Omega^2-k_z^2}\pm \frac{|k_z|}{\Omega}\sqrt{ \left(1-\frac{(k_{-}^{f})^2}{\Omega^2-k_z^2} \right)\left( 1-\frac{(k_{+}^{f})^2}{\Omega^2-k_z^2}\right)} 
\right], \label{E1-solution}
\end{equation}
\begin{equation}
 \left. E_{n^{\prime},p_z-k_z,f} \right|_{p_z = p_{z,f}^{(\pm)} } =\frac{\lambda \eta \Omega}{2} \left[
1-\frac{2(n-n^{\prime})|e_fB|}{\Omega^2-k_z^2}\mp \frac{|k_z|}{\Omega}\sqrt{ \left(1-\frac{(k_{-}^{f})^2}{\Omega^2-k_z^2} \right)\left( 1-\frac{(k_{+}^{f})^2}{\Omega^2-k_z^2}\right)} 
\right].
\label{E2-solution}
\end{equation}
\end{subequations}
By making use of these solutions, we can easily perform the integration over the longitudinal momentum  $p_z$ in Eq.~(\ref{eqE7}). The result reads
\begin{eqnarray}
\mbox{Im}\Pi^{(i)} (\Omega;\mathbf{k}) &=& \sum_{f=u,d}\frac{N_c \alpha_f }{4\pi \ell_f^4} \sum_{n,n^\prime = 0}^{\infty} \sum_{s,s^\prime,\eta,\lambda=\pm 1} \Theta_{\lambda, \eta}^{n,n^{\prime}}(\Omega,k_z) \mathcal{F}_{s,f}^{(i)}(\xi,\Omega) \nonumber\\
&\times&
\frac{ n_F(E_{n,p_z,f}+\eta \mu+\eta s \mu_5)-n_F(\lambda E_{n^\prime,p_z-k_z,f}+\eta \mu+\eta s \mu_5) }{\eta\lambda  \sqrt{ \left( \Omega^2-k_z^2-(k_{-}^f)^2 \right) \left( \Omega^2-k_z^2-(k_{+}^f)^2\right)}} \Bigg|_{p_z = p_{z,f}^{(s^\prime)}},
\end{eqnarray}
where the threshold function $\Theta_{\lambda, \eta}^{n,n^{\prime}}(\Omega,k_z)$ is defined as follows:
\begin{equation}
\Theta_{\lambda, \eta}^{n,n^{\prime}}(\Omega,k_z) = \left\{
\begin{array}{lll}
  \theta\left((k_{-}^f)^2 + k_z^2 - \Omega^2 \right) & \mbox{for}& \lambda=1, \eta=-1, n>n^{\prime},\\
  \theta\left( (k_{-}^f)^2 + k_z^2 - \Omega^2 \right) & \mbox{for}& \lambda=1, \eta=1, n<n^{\prime},\\
  \theta\left( \Omega^2 - k_z^2 -(k_{+}^f)^2 \right) & \mbox{for}& \lambda=-1, \eta=-1,
  \end{array}
\right.
\end{equation} 
and $\Theta_{\lambda, \eta}^{n,n^{\prime}}(\Omega,k_z) =0$ otherwise. By definition, $\theta(x)$ is the Heaviside step function. 

The explicit expressions for functions in Eqs.~(\ref{F-0}) -- (\ref{F-par}) on the solutions to the energy-momentum relation read
\begin{eqnarray}
 \left.\mathcal{F}_{s,f}^{(0)}(\xi,\Omega)\right|_{p_z = p_{z,f}^{(s^\prime)} } &=& \mathcal{A}_{s,s^\prime,f}(\xi,\Omega) + \mathcal{B}_{f}(\xi,\Omega), 
 \label{F-0-s-prime}  \\
\left.\mathcal{F}_{s,f}^{(-)}(\xi,\Omega)\right|_{p_z = p_{z,f}^{(s^\prime)} } &=& \pi \ell_f^2 
\left( \Omega^2-k_z^2 \right)
\left[ 1 - \frac{2(n+n^\prime)}{\left( \Omega^2-k_z^2 \right)\ell_f^2 }- s s^\prime \frac{k_z}{|k_z|}
\sqrt{\left(1-\frac{(k_{-}^{f})^2}{\Omega^2-k_z^2} \right) \left( 1-\frac{(k_{+}^{f})^2}{\Omega^2-k_z^2}\right)}\right]
\nonumber\\
&&\times  \left[(1+s_\perp)\mathcal{I}_{0,f}^{n-1,n^\prime}(\xi)+(1-s_\perp)\mathcal{I}_{0,f}^{n,n^\prime-1}(\xi)\right],
 \label{F-minus-s-prime}\\
\left.\mathcal{F}_{s,f}^{(+)}(\xi,\Omega)\right|_{p_z = p_{z,f}^{(s^\prime)} } &=&  \pi \ell_f^2 
\left( \Omega^2-k_z^2 \right)
\left[ 1 - \frac{2(n+n^\prime) }{\left( \Omega^2-k_z^2 \right)\ell_f^2 } + s s^\prime \frac{k_z}{|k_z|}
\sqrt{\left(1-\frac{(k_{-}^{f})^2}{\Omega^2-k_z^2} \right) \left( 1-\frac{(k_{+}^{f})^2}{\Omega^2-k_z^2}\right)}\right]  \nonumber\\
&&\times \left[(1-s_\perp)\mathcal{I}_{0,f}^{n,n^\prime-1}(\xi)+(1-s_\perp)\mathcal{I}_{0,f}^{n-1,n^\prime}(\xi)\right], 
\label{F-plus-s-prime}\\
 \left.\mathcal{F}_{s,f}^{(\parallel)}(\xi,\Omega)\right|_{p_z = p_{z,f}^{(s^\prime)} } &=& -\mathcal{A}_{s,s^\prime,f}(\xi,\Omega)+ \mathcal{B}_{f}(\xi,\Omega).
  \label{F-par-s-prime}
\end{eqnarray}
Here, we used the following two auxiliary functions: 
\begin{eqnarray}
\mathcal{A}_{s,s^\prime,f}(\xi,\Omega) &=& 2\pi\frac{\Omega^2+k_z^2 }{\Omega^2-k_z^2} \Bigg\{  \frac{2(n-n^\prime)^2 }{\left( \Omega^2-k_z^2 \right)\ell_f^2 }- (n+n^\prime) + \frac{2s^\prime\Omega  |k_z| (n-n^{\prime})}{\Omega^2+k_z^2}
 \nonumber\\
 &\times& \sqrt{\left(1-\frac{(k_{-}^{f})^2}{\Omega^2-k_z^2} \right)
\left( 1-\frac{(k_{+}^{f})^2}{\Omega^2-k_z^2}\right)} \Bigg\}\left[\mathcal{I}_{0,f}^{n,n^\prime}(\xi)+\mathcal{I}_{0,f}^{n-1,n^\prime-1}(\xi)\right]\nonumber\\
&+&\frac{4\pi s_\perp s  \Omega k_z }{\Omega^2-k_z^2} \Bigg\{  \frac{2(n-n^\prime)^2 }{\left( \Omega^2-k_z^2 \right)\ell_f^2 }- (n+n^\prime) 
+\frac{s^\prime(n-n^{\prime}) (\Omega^2+k_z^2)}{2\Omega |k_z|} 
 \nonumber\\
 &\times& \sqrt{\left(1-\frac{(k_{-}^{f})^2}{\Omega^2-k_z^2} \right)
\left( 1-\frac{(k_{+}^{f})^2}{\Omega^2-k_z^2}\right)} \Bigg\}\left[\mathcal{I}_{0,f}^{n,n^\prime}(\xi)-\mathcal{I}_{0,f}^{n-1,n^\prime-1}(\xi)\right] ,\\
\mathcal{B}_{f}(\xi,\Omega) &=&4\pi\left\{(n+n^\prime)\left[\mathcal{I}_{0,f}^{n-1,n^\prime-1}(\xi)+\mathcal{I}_{0,f}^{n,n^\prime}(\xi)\right]-\xi\left[\mathcal{I}_{0,f}^{n,n^\prime-1}(\xi)+\mathcal{I}_{0,f}^{n-1,n^\prime}(\xi)\right]\right\}.
\end{eqnarray}
Note that  $\mathcal{B}_{f}(\xi,\Omega)$ is independent of $s$ and $s^\prime$.


%

\end{document}